\documentclass[12pt,english]{iopart}
\pdfoutput=1
\usepackage[T1]{fontenc}
\usepackage[latin9]{inputenc}
\usepackage{amstext}
\usepackage{amssymb}
\usepackage{graphicx}

\makeatletter

\newcommand{\lyxdot}{.}

\usepackage{iopams}

\@ifundefined{showcaptionsetup}{}{%
 \PassOptionsToPackage{caption=false}{subfig}}
\usepackage{subfig}
\makeatother

\usepackage{babel}
\begin{document}

\title{Strong disorder renormalization group study of aperiodic quantum
Ising chains}

\author{Fleury J. Oliveira Filho, Maicon S. Faria, and André P. Vieira}

\address{Instituto de Física da USP, São Paulo, SP, Brazil}
\begin{abstract}
We employ an adaptation of a strong-disorder renormalization-group
technique in order to analyze the ferro-paramagnetic quantum phase
transition of Ising chains with aperiodic but deterministic couplings
under the action of a transverse field. In the presence of marginal
or relevant geometric fluctuations induced by aperiodicity, for which
the critical behavior is expected to depart from the Onsager universality
class, we derive analytical and asymptotically exact expressions for
various critical exponents (including the correlation-length and the
magnetization exponents, which are not easily obtainable by other
methods), and shed light onto the nature of the ground state structures
in the neighborhood of the critical point. The main results obtained
by this approach are confirmed by finite-size scaling analyses of
numerical calculations based on the free-fermion method.
\end{abstract}
\maketitle

\section{Introduction}

Quantum phase transitions differ from thermal transitions in that
they happen at zero temperature, induced by quantum fluctuations ultimately
inherent to all physical systems. As these transitions occur at zero
temperature, they can be verified even in one-dimensional systems,
which are especially sensitive to the effects of both quantum and
geometric fluctuations, being thus ideal laboratories for the study
of exotic phenomena associated with the presence of disorder. 

Arguably the simplest spin model to undergo a quantum phase transition
is the quantum Ising chain, described by the Hamiltonian
\begin{equation}
H=-\frac{1}{2}\sum_{j=1}^{N}J_{j}\sigma_{j}^{x}\sigma_{j+1}^{x}-\frac{1}{2}\sum_{j=1}^{N}h_{j}\sigma_{j}^{z},\label{eq:hamiltIsingchain}
\end{equation}
in which the $\sigma^{x,z}$ are Pauli matrices, $J_{j}>0$ represents
a ferromagnetic bond between the longitudinal components of the spins
at sites $j$ and $j+1$, and $h_{j}$ denotes a transverse field
acting on the spin at site $j$. In the uniform limit ($J_{j}\equiv J$,
$h_{j}\equiv h>0$), there is a quantum phase transition at $h=J$,
between a ferromagnetic phase dominated by the nearest-neighbor bonds
$J$, and a paramagnetic phase, dominated by the fields $h$. In this
case, the model corresponds to the extreme anisotropy limit of the
two-dimensional Ising model. Indeed, the quantum transition at $h=J$
belongs to the Onsager universality class, whose paradigm is precisely
the thermal transition of the two-dimensional Ising model.

In the presence of disorder represented by randomly chosen couplings
$J_{j}$ and $h_{j}$, the model in Eq. \ref{eq:hamiltIsingchain}
exhibits surprising behavior. By extending a strong-disorder real-space
renormalization group (SDRG) scheme introduced by Ma, Dasgupta, and
Hu \cite{ma79,dasgupta80} to study the random Heisenberg chain, Fisher
\cite{fisher92,fisher95} was able to show that the nature of the
quantum transition is fundamentally changed with respect to that of
the uniform limit, being now characterized by a brutal distinction
between average and typical behavior of various physical quantities,
such as correlation functions. At criticality, there is a modification
of the dynamic relation between length and time scales, which assumes
an activated form (\emph{à la} Arrhenius), in contrast with the usual
power law. Sufficiently close to criticality, there are Griffiths
phases \cite{griffiths69}, in which some thermodynamic properties
are still singular. Fisher also provided exact calculations of a series
of other properties for the random system, many of which are not known
in the uniform limit, such as the scaling form for the magnetization
as a function of an ordering field at the critical point. 

Partially similar effects are verified in the presence of aperiodic
but deterministic couplings. This kind of aperiodicity is inspired
by analogy with quasicrystals \cite{schechtman84}, which exhibit
symmetries forbidden by traditional crystallography, corresponding
to projections of high-dimensional Bravais lattices on low-dimensional
subspaces. In one spatial dimension, aperiodic couplings can be defined
from sequences of letters obtained from substitution or inflation
rules, such as the one associated with the Fibonacci sequence, $a\rightarrow ab$,
$b\rightarrow a$. Iteration of this rule yields a sequence $abaababa\ldots$
with no characteristic period. By associating different letters with
different bond values $J_{a}$ and $J_{b}$ (or with different field
values $h_{a}$ and $h_{b}$) one obtains an aperiodic quantum Ising
chain. Any aperiodic sequence induces geometric fluctuations gauged
by a wandering exponent $\omega$, which describes the growth of these
fluctuations as the chain length increases. A value of $\omega=\frac{1}{2}$
would mimic the geometric fluctuations induced by random couplings.

For aperiodic quantum Ising chains, there are exact results obtained
by RG treatments, unrelated to the SDRG scheme, and valid at the critical
point \cite{luck93a,igloi97b,hermisson97}. These results allow the
inference of the relation between length and time scales, and show
that aperiodic sequences characterized by a wandering exponent $\omega=\frac{1}{2}$
induce, at the critical point, effects analogous to those produced
by randomness. Additionally, they confirm the validity of a heuristic
criterion formulated by Luck \cite{luck93b} in order to gauge the
effects of aperiodicity on ferromagnetic phase transitions. According
to this criterion, aperiodicity is relevant, \emph{i.e.} it is capable
of changing the critical behavior of the quantum Ising chain, whenever
it is associated with a wandering exponent $\omega>0$. The marginal
case $\omega=0$ may lead to nonuniversal critical behavior, dependent
on the coupling values.

In the present paper, we adapt the SDRG scheme to study the critical
behavior of quantum Ising chains with aperiodic couplings. We show
that this adaptation allows us to investigate the system both at the
critical point and in its neighborhood, and to obtain presumably exact
analytical results for a series of critical exponents, such as those
associated with the bulk magnetization, the critical correlations,
and the correlation length, which are not easily accessible via other
methods. This allows a comparison with numerical results obtained
for some of these exponents by Iglói \emph{et al.} from a mapping
of the quantum Ising chain to a free-fermion problem \cite{igloi98,igloi98b}.
We note that essentially the same SDRG approach has been employed
to study the scaling form of the entanglement entropy in aperiodic
quantum Ising chains, although only at the critical point \cite{szallas2005,igloi2007}.

This paper is organized as follows. In Section \ref{sec:Random-quantum-Ising-chains-and-the-SDRG-approach}
we review the SDRG approach introduced to deal with the random quantum
Ising chain, as well as the associated low-energy physics. In Section
\ref{sec:Aperiodic-sequences-and-the-Harris-Luck-criterion} we introduce
the kind of aperiodic structures in which we focus on this paper,
as well as the notion of geometric fluctuations, discussing their
connection with the Harris-Luck criterion for the relevance of aperiodicity
to the critical behavior of the quantum Ising chain. Section \ref{sec:A-family-of-aperiodic-sequences}
presents our extension of the SDRG approach to deal with aperiodic
quantum Ising chains with couplings chosen from a family of aperiodic
sequences whose geometric fluctuations can be tuned from irrelevant
to relevant. In Section \ref{sec:Rudin-Shapiro-sequences} we discuss
the behavior of quantum Ising chains with couplings following Rudin-Shapiro
sequences, whose geometric fluctuations mimic those associated with
random couplings. The final section presents our concluding remarks.

\section{Random quantum Ising chains and the strong-disorder RG approach}

\label{sec:Random-quantum-Ising-chains-and-the-SDRG-approach}The
random quantum Ising chain corresponds to the extremely anisotropic
limit of the McCoy-Wu model \cite{mccoy68,mccoy71}, and its very
interesting low-temperature properties have been elucidated by Fisher
\cite{fisher92,fisher95}, who argued that an adaptation of the strong-disorder
renormalization group (SDRG) approach of Ma, Dasgupta and Hu \cite{ma79,dasgupta80}
yields asymptotically exact results for various physical quantities.
(For a review of the SDRG method, as applied to the random quantum Ising chain
and to many other problems, see Ref. \cite{igloimonthus2005}.)

The basic idea of the SDRG scheme is the iterative elimination of
high-energy degrees of freedom, associated with locally strong couplings.
For sufficiently strong randomness, in which there are very few strong
couplings and many weak couplings, this is easily defined. In a finite
chain, there is always a strongest coupling. If this coupling is a
field, say $h_{2}$, most likely the neighboring bonds $J_{1}$ and
$J_{2}$ are much weaker, and the spin $\sigma_{2}$ will tend to
align with the transverse field, giving negligible contributions to
magnetic properties. If we are interested in the low-energy behavior,
the spin $\sigma_{2}$, along with the field $h_{2}$ and the bonds
$J_{1}$ and $J_{2}$, can be eliminated from the system. Nevertheless,
virtual excitations will induce an effective coupling between the
$x$ components of the neighboring spins $\sigma_{1}$ and $\sigma_{3}$,
given, up to second order in perturbation theory, by
\begin{equation}
\tilde{J}_{1}=\frac{J_{1}J_{2}}{h_{2}}.\label{eq:jeffrqic}
\end{equation}
If the strongest coupling is a bond, say $J_{2}$, the spins $\sigma_{1}$
and $\sigma_{2}$ that it connects are most likely under the action
of much weaker fields $h_{1}$ and $h_{2}$, and thus will tend to
respond, at low energies, as a single effective spin, under the action
of an effective transverse field created by virtual excitations and
given, again up to second order, by
\begin{equation}
\tilde{h}_{1}=\frac{h_{1}h_{2}}{J_{2}}.\label{eq:heffrqic}
\end{equation}
The bond $J_{2}$ and the fields $h_{1}$ and $h_{2}$, along with
the spins $\sigma_{1}$ and $\sigma_{2}$, are then replaced by the
effective spin, with an effective magnetic moment, acted on by the
field $\tilde{h}_{1}$.

In any case, the overall energy scale is reduced, since 
both effective couplings $\tilde{h}_{1}$ and $\tilde{J}_{1}$ are 
always weaker than any of 
the original couplings. As the transformations described above
are iterated, this leads to a renormalization of the probability distributions of
both bonds and fields, as well as lengths and magnetic moments. The
fixed-point distributions can be found in closed form \cite{fisher95,igloi2002},
and from these many low-energy physical properties can be derived.
The nature of the ground state is strongly modified by the presence
of disorder. While there is still a well-defined critical point, which
corresponds to Pfeuty's exact result \cite{pfeuty79},
\begin{equation}
\overline{\ln h_{i}}=\overline{\ln J_{i}},
\end{equation}
the critical relation between length ($l$) and energy ($\Omega$)
scales departs from the power law valid in the uniform chain, having
now the activated form
\begin{equation}
\Omega\sim\exp\left(-\sqrt{l/l_{0}}\right), 
\end{equation}
where $l_{0}$ is some nonuniversal length scale. Most strikingly,
in the neighborhood of the critical point the model is still gapless
and the dynamic scaling behavior follows
\begin{equation}
\Omega\sim l^{-z\left(\delta\right)}, 
\end{equation}
in which $z\left(\delta\right)$ is a dynamical exponent dependent
on the distance to the critical point, measured by 
\begin{equation}
\delta\sim\overline{\ln J_{i}}-\overline{\ln h_{i}}.
\label{eq:deltarandom}
\end{equation}
This gapless noncritical regime arises from the existence of rare
but arbitrarily large regions of spins which, due to statistical fluctuations,
are locally in the {}``wrong'' phase as compared to the whole chain,
yielding the so-called Griffiths singularities \cite{griffiths69}. 

Interestingly, the above results seem to be asymptotically valid not
only for strong disorder, but for any amount of randomness present
in the system. The reason is most easily seen at criticality, where
the probability distributions of effective couplings become more and
more singular around the origin, rendering the perturbative expressions
for the effective couplings essentially exact. In the following sections,
we show that also in the presence of relevant aperiodic modulations
it is possible to obtain asymptotically exact results, although the
nature of the ground state turns out to be quite distinct from that
of the random case.

\section{Aperiodic sequences and the Harris-Luck criterion}

\label{sec:Aperiodic-sequences-and-the-Harris-Luck-criterion}Substitution
rules defining aperiodic sequences have the general form
\[
\left\{ a_{i}\rightarrow w_{i}\right.,\quad i\in\left\{ 1,2,\dots,n\right\} ,
\]
in which $w_{i}$ is a {}``word'' formed by letters in the set $\left\{ a_{i}\right\} $.
The sequence is built by iteratively applying the rule to an initial
letter. For instance, the celebrated Fibonacci sequence is obtained
by applying the rule
\[
\left\{ \begin{array}{l}
a\rightarrow ab\\
b\rightarrow a
\end{array}\right.
\]
to an initial letter $a$. 

Various properties of the resulting infinite sequence (obtained after
an infinite number of iterations of the rule) are related to the substitution
matrix
\[
\mathbb{M}=\left(\begin{array}{cccc}
f_{11} & f_{12} & \cdots & f_{1n}\\
f_{21} & f_{22} & \cdots & f_{2n}\\
\vdots & \vdots & \ddots & \vdots\\
f_{n1} & f_{n2} & \cdots & f_{nn}
\end{array}\right),
\]
in which $f_{ij}$ denotes the number of letters $a_{i}$ in the word
$w_{j}$. For instance, the length of the sequence after $k$ iterations
of the substitution rule is asymptotically given by
\[
N_{k}\sim\zeta_{1}^{k},
\]
$\zeta_{1}>0$ being the largest eigenvalue of $\mathbb{M}$. Also,
the $i$th entry of the corresponding eigenvector (normalized so as
the sum of all entries equals unity) gives the relative frequency
$d_{i}$ of the different letters in the infinite sequence. 

For the purposes of the present paper, the most relevant property
derived from the substitution matrix are the fluctuations of the relative
frequency of a letter, after $k$ iterations of the rule, with respect
to that of the infinite sequence. These fluctuations are explicitly
defined as 
\[
g_{k}^{\left(i\right)}=\left|N_{k}^{\left(i\right)}-d_{i}N_{k}\right|,
\]
where $N_{k}^{\left(i\right)}$ denotes the number of letters $a_{i}$
after $k$ iterations of the substitution rule. It can be shown that
$g_{k}^{\left(i\right)}$, for all $i$, is asymptotically governed
by the second largest eigenvalue (in absolute value) of $\mathbb{M}$,
\[
g_{k}^{\left(i\right)}\sim\left|\zeta_{2}\right|^{k}\equiv N_{k} ^{\omega},
\]
in which
\[
\omega\equiv\frac{\ln\left|\zeta_{2}\right|}{\ln\zeta_{1}}
\]
is the so-called wandering exponent of the aperiodic sequence. A negative
value of $\omega$ indicates that fluctuations are bounded, and become
negligible as the length of the sequence becomes larger and larger.
On the other hand, a positive $\omega$ leads to unbounded fluctuations,
as in the case of a sequence built from independently and randomly
distributed letters, for which $\omega=\frac{1}{2}$. Finally, the
marginal case of $\omega=0$ usually leads to fluctuations which grow
logarithmically with the sequence length. 

When couplings in a ferromagnetic model are chosen in a nonperiodic
fashion, the Harris-Luck heuristic criterion \cite{luck93b,harris74}
makes use of the concept of the fluctuations of those couplings in
a length scale defined by the correlation length to point out under
which conditions aperiodicity changes the critical behavior with respect
to that of the uniform system. In the case of a quantum Ising chain
with bonds chosen from some aperiodic sequence, such fluctuations
are proportional to the $g_{k}^{\left(i\right)}$ defined above, and
the criterion states that aperiodicity will be relevant, \emph{i.e.}
the critical behavior will be changed, whenever the wandering exponent
satisfies
\begin{equation}
\omega\geq1-\frac{1}{\nu}=0,\label{eq:hlc}
\end{equation}
in which $\nu=1$ is the correlation-length critical exponent of the
Onsager universality class.

In the following section, we analyze the effects of bonds chosen from
a family of aperiodic sequences with wandering exponents ranging from
$\omega=-1$ to $\omega\rightarrow1$.

\section{A family of aperiodic sequences}

\label{sec:A-family-of-aperiodic-sequences}
\begin{figure}
\includegraphics[width=0.99\textwidth]{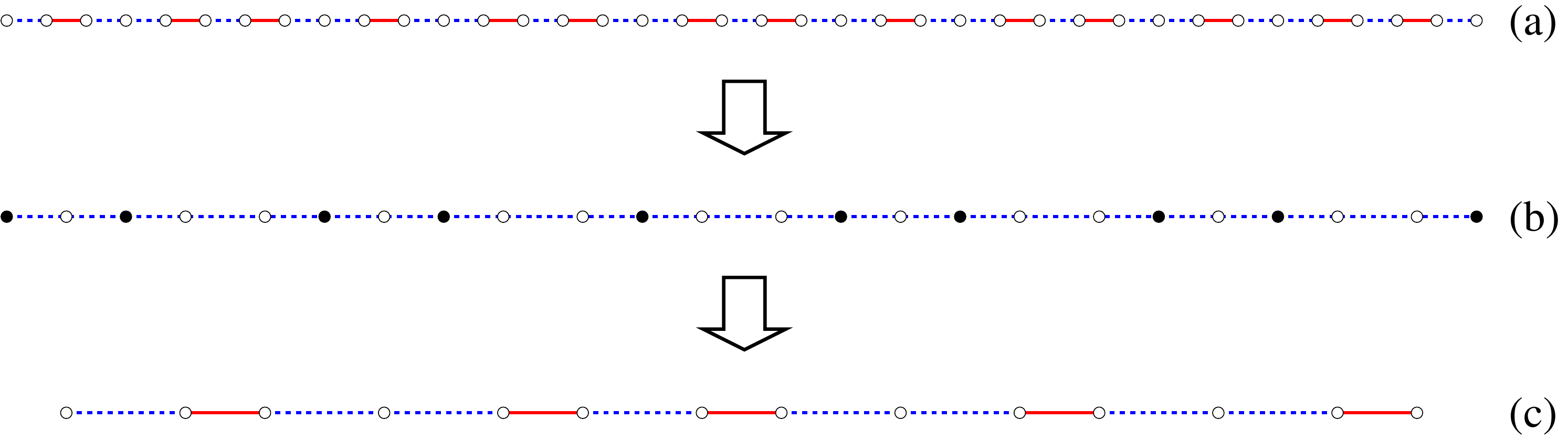}

\caption{\label{fig:fib}Renormalization-group transformation as applied to
the Fibonacci chain. The original chain is depicted in (a), with weak
$J_{a}$ bonds indicated by dashed (blue) lines and strong $J_{b}$
bonds by solid (red) lines; spins sit on the open circles, under the
action of the original transverse field $h$. After the spins connected
by strong bonds are joined, the chain takes the form shown in (b),
with open circles now denoting effective fields $\tilde{h}$ and
solid circles representing original fields $h$. Decimation of the
spins under the action of $h$ leads to the sequence of effective
couplings shown in (c), which again corresponds to a Fibonacci chain.}
\end{figure}
Consider the family of aperiodic sequences generated by the substitution
rule \cite{grimm1997}
\begin{equation}
\left\{ \begin{array}{l}
a\rightarrow ab^{k}\\
b\rightarrow a
\end{array}\right.,\qquad b^{k}\equiv\underbrace{bb\cdots b}_{k\ \mathrm{letters}}\label{eq:substitutionrule}
\end{equation}
with $k$ a positive integer, so that $k=1$ corresponds to the Fibonacci
sequence, illustrated in Fig. \ref{fig:fib}(a). The substitution
matrix has eigenvalues
\[
\zeta_{k}^{\pm}=\frac{1}{2}\pm\frac{1}{2}\sqrt{1+4k},
\]
and thus the wandering exponent,
\begin{equation}
\omega_{k}=\frac{\ln\left|\zeta_{k}^{-}\right|}{\ln\zeta_{k}^{+}}=\frac{\ln k}{\ln\zeta_{k}^{+}}-1,
\end{equation}
is a monotonically increasing function of $k$, with $\omega_{1}=-1$,
$\omega_{2}=0$, $\omega_{3}=0.317\dots$, etc. According to the Harris-Luck
criterion, we thus expect that aperiodicity is relevant for $k\geq3$,
with $k=2$ representing a marginal case.

The sequences generated by Eq. \ref{eq:substitutionrule} contain
clusters of $k$ letters $b$, separated by clusters of $1$ or $k+1$
letters $a$. In order to implement a strong-disorder RG treatment
of the corresponding quantum Ising chain, we have to calculate the
effective bonds and fields associated with those clusters. Although
we start with a uniform field, the RG scheme will generate different
effective fields for different clusters, and this will be taken into
account in the discussion below.

\begin{figure}
\centering{}\subfloat[]{\centering{}\includegraphics[width=0.4\columnwidth]{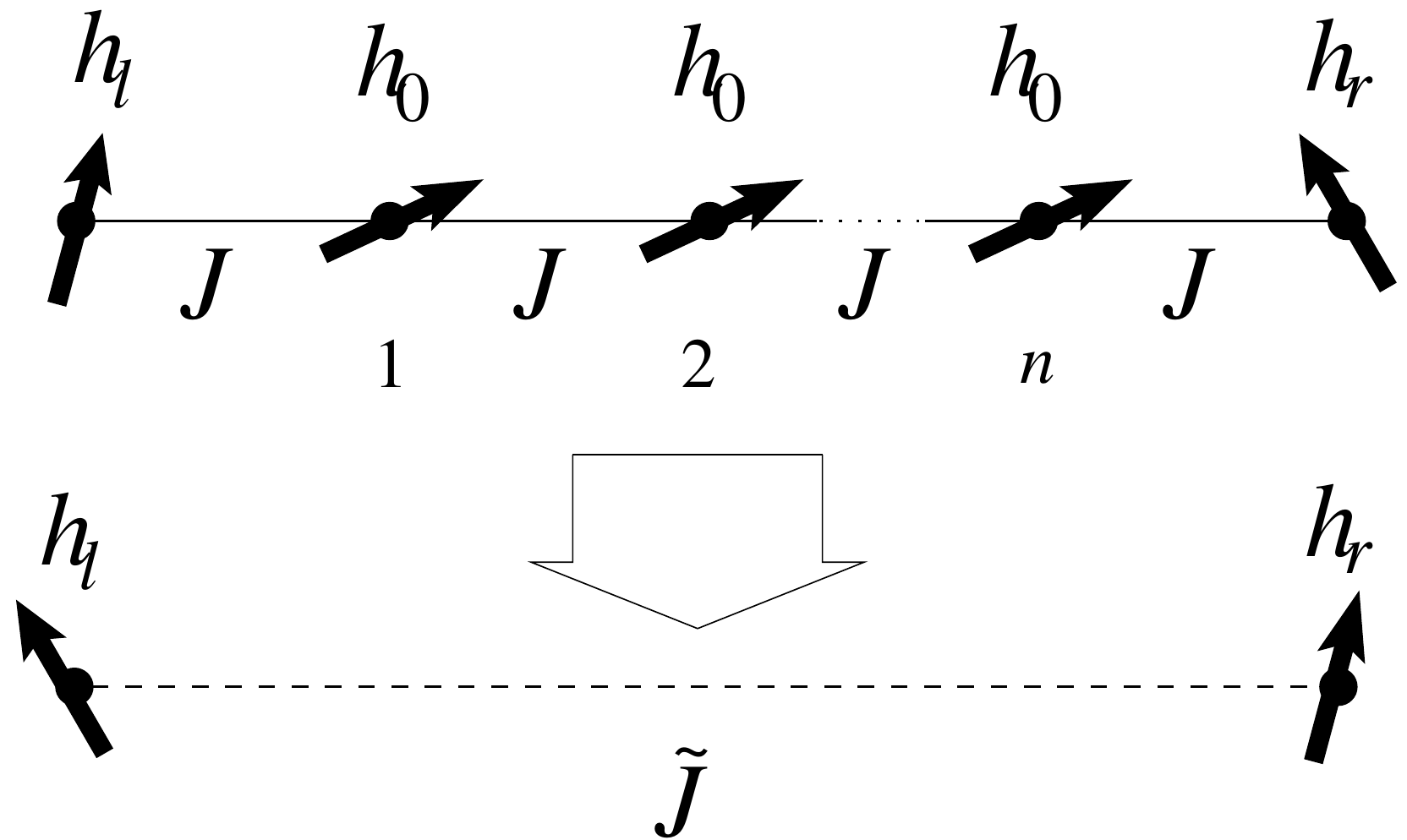}}\hfill{}
\subfloat[]{\centering{}\includegraphics[width=0.4\columnwidth]{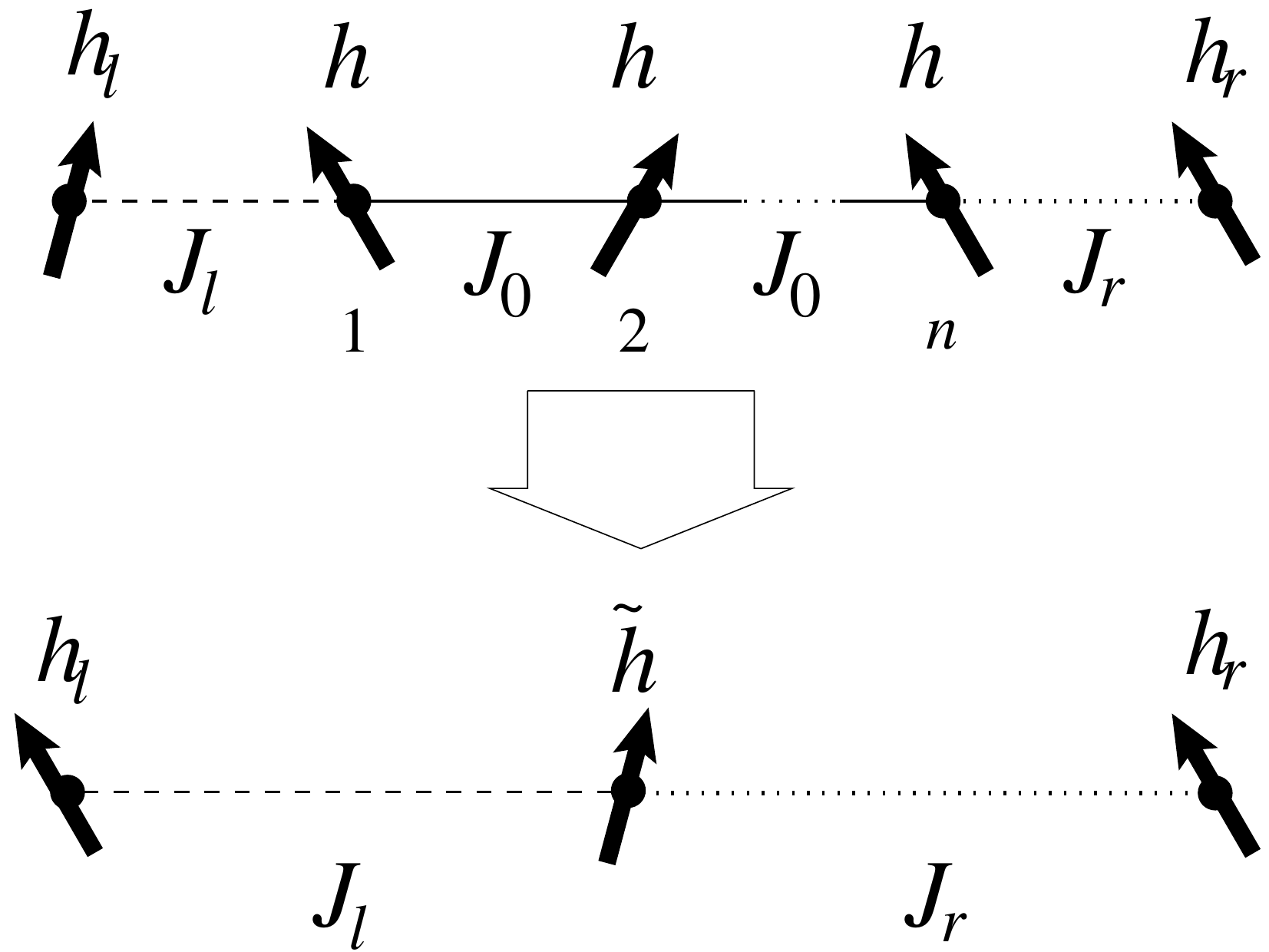}}\caption{Renormalization of couplings and decimation of spins in the aperiodic
quantum Ising chain. (a) Elimination of spins connected by $n$ neighboring
strong fields $h_{0}$. (b) Formation of a effective spin from a cluster
of $n$ spins connected by strong bonds $J_{0}$.\label{fig:Renormalization-of-couplings}}
\end{figure}

We first look at the case of $n$ spins $\sigma_{i}$ ($i=1,2,\dots,n$)
under the action of a field $h_{0}$, forming with spins $\sigma_{l}$
and $\sigma_{r}$ a cluster connected by $n+1$ bonds $J\ll h_{0}$;
see Fig. \ref{fig:Renormalization-of-couplings}(a). Furthermore,
suppose that the fields $h_{l}$ and $h_{r}$ acting on $\sigma_{l}$
and $\sigma_{r}$ are small compared to $h_{0}$. Physically, one
expects that the $n$ spins in the cluster tend to line up with the
field $h_{0}$, and thus can be decimated out of the system. Applying
perturbation theory up to order $n+1$ leads to the conclusion that
fluctuations of the cluster induce an effective coupling
\begin{equation}
\tilde{J}=\frac{J^{n+1}}{h_{0}^{n}}\label{eq:jeffn}
\end{equation}
between $\sigma_{l}$ and $\sigma_{r}$. 

A more delicate situation corresponds to a cluster of $n$ spins connected
by strong bonds $J_{0}$, while the field $h$ acting on each of the
$n$ spins, as well as the bonds $J_{l}$ and $J_{r}$ to the rest
of the chain, are much smaller; see Fig. \ref{fig:Renormalization-of-couplings}(b).
However, this can be easily dealt with by resorting to a duality transformation
\cite{kogut79}, which interchanges the roles of bonds and fields.
In the transformed space, the situation is precisely the one of the
previous paragraph, so that the result can be immediately written
down. At low energies, the $n$ spins behave as a single magnetic
moment, under the action of an effective field
\begin{equation}
\tilde{h}=\frac{h^{n}}{J_{0}^{n-1}},\label{eq:heffn}
\end{equation}
with an effective magnetic moment
\begin{equation}
\tilde{\mu}=n\mu,\label{eq:mueffn}
\end{equation}
$\mu$ being the magnetic moment of each of the $n$ spins.

Getting back to the sequences generated by Eq. \ref{eq:substitutionrule},
suppose we look at intermediate values of the field, so that $J_{a}\ll h\ll J_{b}$.
(The case $J_{a}\gg J_{b}$ leads to a renormalized chain in which
the roles of weak and strong bonds are interchanged, as long as the
system is sufficiently close to criticality. Thus, there is no loss
of generality in assuming $J_{b}\gg J_{a}$.) At energy scales $\Omega$
such that $h<\Omega<J_{b}$, Eqs. \ref{eq:heffn} and \ref{eq:mueffn}
imply that each cluster of $k+1$ spins connected by bonds $J_{b}$
behaves as a single magnetic moment, under the action of an effective
field
\begin{equation}
\tilde{h}=\frac{h^{k+1}}{J_{b}^{k}},\label{eq:heffk}
\end{equation}
with an effective magnetic moment
\begin{equation}
\tilde{\mu}=\left(k+1\right)\mu.\label{eq:mueffk}
\end{equation}
As $\tilde{h}<h$, and $J_{a}<h$, the largest coupling is now provided
by the original field $h$, as illustrated in Fig. \ref{fig:fib}(b)
for the case of the Fibonacci chain. Further reducing the energy scale,
so that $J_{a}<\Omega<h$, the clusters of $k+1$ bonds $J_{a}$,
containing $k$ spins under the action of the field $h$, can be decimated
out, leaving effective bonds
\begin{equation}
\tilde{J}_{a}=\frac{J_{a}^{k+1}}{h^{k}},\label{eq:jaeffk}
\end{equation}
as shown in Fig. \ref{fig:fib}(c). All original fields $h$ are decimated
in this step, so that all active spins at this energy scale are under
the action of the same field $\tilde{h}$. Now the largest couplings
in the chain are provided by the remaining clusters of isolated $J_{a}$
bonds, which we can interpret as effective bonds
\begin{equation}
\tilde{J}_{b}=J_{a},\label{eq:jbeffk}
\end{equation}
since with this choice the sequence of effective bonds exactly reproduces
the original sequence. Equations \ref{eq:heffk} through \ref{eq:jbeffk}
define an RG step.

In general, there is no guarantee that the effective couplings satisfy
the condition $\tilde{J}_{a}<\tilde{h}<\tilde{J}_{b}$, which is required
to iterate the RG procedure. However, we shall assume that the condition
is satisfied, and explore the consequences. We first introduce a change
in notation, rewriting the RG equations as
\begin{equation}
J_{a}^{(j+1)}=\frac{\left[J_{a}^{(j)}\right]^{k+1}}{\left[h^{(j)}\right]^{k}},\qquad J_{b}^{(j+1)}=J_{a}^{(j)},\label{eq:jajbeff}
\end{equation}
\begin{equation}
h^{(j+1)}=\frac{\left[h^{(j)}\right]^{k+1}}{\left[J_{b}^{(j)}\right]^{k}},\qquad\mu^{(j+1)}=\left(k+1\right)\mu^{(j)},\label{eq:hmueff}
\end{equation}
in which $J_{a}^{(j)}$, $J_{b}^{(j)}$, $h^{(j)}$ and $\mu^{(j)}$
represent the values of the parameters after $j$ iterations of the
RG procedure, with $J_{a}^{(0)}\equiv J_{a}$, $J_{b}^{(0)}\equiv J_{b}$,
$h^{(0)}\equiv h$, and $\mu^{(0)}=1$. Next, we define the ratios
\begin{equation}
r^{(j)}=\frac{h^{(j)}}{J_{b}^{(j)}}\qquad\mbox{and}\qquad s^{(j)}=\frac{J_{a}^{(j)}}{h^{(j)}},\label{eq:rseff}
\end{equation}
which both remain smaller than unity as long as the condition $J_{a}^{(j)}<h^{(j)}<J_{b}^{(j)}$
is fullfilled. From Eqs. \ref{eq:jajbeff} and \ref{eq:hmueff} we
obtain recursion relations for the ratios,
\[
r^{(j+1)}=\frac{\left[r^{(j)}\right]^{k}}{s^{(j)}}\qquad\mbox{and}\qquad s^{(j+1)}=\frac{\left[s^{(j)}\right]^{k+1}}{\left[r^{(j)}\right]^{k}},
\]
which can be written in matrix form by taking the logarithm of the
previous equations, yielding
\begin{equation}
\left(\begin{array}{c}
\ln r^{(j+1)}\\
\ln s^{(j+1)}
\end{array}\right)=\left(\begin{array}{cc}
k & -1\\
-k & k+1
\end{array}\right)\left(\begin{array}{c}
\ln r^{(j)}\\
\ln s^{(j)}
\end{array}\right).\label{eq:lnr-recrel}
\end{equation}
Defining
\begin{equation}
\left|v_{j}\right\rangle =\left(\begin{array}{c}
\ln r^{(j)}\\
\ln s^{(j)}
\end{array}\right)\qquad\mbox{and}\qquad\mathbb{T}_{k}=\left(\begin{array}{cc}
k & -1\\
-k & k+1
\end{array}\right),\label{eq:vjTk}
\end{equation}
and expressing the matrix $\mathbb{T}_{k}$ as $\mathbb{T}_{k}=\mathbb{U}_{k}\mathbb{D}_{k}\mathbb{U}_{k}^{-1}$,
in which the diagonal matrix $\mathbb{D}_{k}$ consists of the eigenvalues
of $\mathbb{T}_{k}$ and the columns of the matrix $\mathbb{U}_{k}$
contain the corresponding right eigenvectors, we can rewrite Eq. \ref{eq:lnr-recrel}
as 
\[
\left|u_{j+1}\right\rangle \equiv\mathbb{U}_{k}^{-1}\left|v_{j+1}\right\rangle =\mathbb{D}_{k}\mathbb{U}_{k}^{-1}\left|v_{j}\right\rangle \equiv\mathbb{D}_{k}\left|u_{j}\right\rangle ,
\]
from which we readily obtain
\begin{equation}
\left|u_{j}\right\rangle =\mathbb{D}_{k}^{j}\left|u_{0}\right\rangle \quad\Rightarrow\quad\left|v_{j}\right\rangle =\mathbb{U}_{k}\mathbb{D}_{k}^{j}\mathbb{U}_{k}^{-1}\left|v_{0}\right\rangle .\label{eq:vjv0}
\end{equation}

The eigenvalues of the matrix $\mathbb{T}_{k}$ are related to the
eigenvalues of the substitution matrix by
\[
\lambda_{k}^{\pm}=k+\zeta_{k}^{\pm}=\left(\zeta_{k}^{\pm}\right)^{2},
\]
while the matrix $\mathbb{U}_{k}$ can be written as
\[
\mathbb{U}_{k}=\left(\begin{array}{cc}
1 & 1\\
-\zeta_{k}^{+} & -\zeta_{k}^{-}
\end{array}\right).
\]
Using the above results, we can invert $\mathbb{U}_{k}$ and expand
Eq. \ref{eq:vjv0} to obtain
\begin{equation}
\ln r^{(j)}=\frac{1}{\zeta_{k}^{+}-\zeta_{k}^{-}}\left[-\left(\lambda_{k}^{+}\right)^{j}\alpha_{k}^{-}+\left(\lambda_{k}^{-}\right)^{j}\alpha_{k}^{+}\right]\label{eq:lnr0}
\end{equation}
and
\begin{equation}
\ln s^{(j)}=\frac{1}{\zeta_{k}^{+}-\zeta_{k}^{-}}\left[\left(\lambda_{k}^{+}\right)^{j}\zeta_{k}^{+}\alpha_{k}^{-}-\left(\lambda_{k}^{-}\right)^{j}\zeta_{k}^{-}\alpha_{k}^{+}\right],\label{eq:lns0}
\end{equation}
with
\[
\alpha_{k}^{\pm}=\zeta_{k}^{\pm}\ln r^{(0)}+\ln s^{(0)}.
\]

We thus see that the behavior of $r^{(j)}$ and $s^{(j)}$ for $j\gg1$
is in general dominated by the largest eigenvalue of 
$\mathbb{T}_{k}$, $\lambda_{k}^{+}>\lambda_{k}^{-}>0$.
If $\alpha_{k}^{-}$, which enters the coefficient of $\left(\lambda_{k}^{+}\right)^{j}$
in both $\ln r^{(j)}$ and $\ln s^{(j)}$, is positive, we asymptotically
have $\ln r^{(j)}\ll0$ and $\ln s^{(j)}\gg0$, and thus the effective
fields are much smaller than the effective bonds. This corresponds
to the ferromagnetic phase. Conversely, if $\alpha_{k}^{-}$ is negative,
the asymptotic behavior corresponds to $\ln r^{(j)}\gg0$ and $\ln s^{(j)}\ll0$,
the effective fields are much larger than the effective bonds, and
the system is in the paramagnetic phase. The critical point must then
correspond to the condition $\alpha_{k}^{-}=0$, which, in terms of
the original parameters, leads to a critical field
\begin{equation}
h_{\mathrm{crit}}=J_{a}^{\frac{1}{1-\zeta_{k}^{-}}}J_{b}^{-\frac{\zeta_{k}^{-}}{1-\zeta_{k}^{-}}}=J_{a}^{d_{a}}J_{b}^{d_{b}},\label{eq:hck}
\end{equation}
in which $d_{a}$ and $d_{b}=1-d_{a}$ can be shown to be the fractions
of letters $a$ and $b$ in the infinite sequence produced by Eq.
\ref{eq:substitutionrule}. This last result agrees with the exact
result derived by Pfeuty \cite{pfeuty79} for the critical condition
of the quantum Ising chain. That the approximate RG equations used
here are able to lead to the exact result in Eq. \ref{eq:hck} must
be related to the fact that the recurrence relations \ref{eq:heffk},
\ref{eq:jaeffk} and \ref{eq:jbeffk} preserve the dual symmetry of
the quantum Ising chain.

It is also useful to determine the effective lengths of the various
couplings as the RG scheme is iterated. We follow Fisher \cite{fisher95}
and assume that fields and bonds share lengths evenly, so that the
initial bonds have lengths $\ell_{a}^{(0)}=\ell_{b}^{(0)}=\frac{1}{2}$
and the initial field has length $\ell_{h}^{(0)}=\frac{1}{2}$. From
Eqs. \ref{eq:jajbeff} and \ref{eq:hmueff}, and by looking at Fig.
\ref{fig:fib}, we see that these lengths satisfy recursion relations
which can be written in matrix form as
\[
\left(\begin{array}{c}
\ell_{a}^{(j+1)}\\
\ell_{b}^{(j+1)}\\
\ell_{h}^{(j+1)}
\end{array}\right)=\left(\begin{array}{ccc}
k+1 & 0 & k\\
1 & 0 & 0\\
0 & k & k+1
\end{array}\right)\left(\begin{array}{c}
\ell_{a}^{(j)}\\
\ell_{b}^{(j)}\\
\ell_{h}^{(j)}
\end{array}\right)\equiv\mathbb{S}_{k}\left(\begin{array}{c}
\ell_{a}^{(j)}\\
\ell_{b}^{(j)}\\
\ell_{h}^{(j)}
\end{array}\right).
\]
Thus, all effective lengths grow asymptotically with the largest eigenvalue
of the matrix $\mathbb{S}_{k}$. It turns out that the eigenvalues
of $\mathbb{S}_{k}$ are $1$, $\lambda_{k}^{-}$ and $\lambda_{k}^{+}$,
so that
\begin{equation}
\ell_{a}^{(j)}\sim\ell_{b}^{(j)}\sim\ell_{h}^{(j)}\sim\left(\lambda_{k}^{+}\right)^{j}.\label{eq:lalblh}
\end{equation}

In the following subsections, we analyze the consequences of the previous
expressions for the behavior of the system both at the critical point
and in its neighborhood.

\subsection{Properties at the critical point}

\label{sub:critical-point}The recursion relations \ref{eq:jajbeff}
and \ref{eq:hmueff} can be iterated indefinitely at the critical
point. As then $\alpha_{k}^{-}=0$, the behavior of the ratios $r^{(j)}$
and $s^{(j)}$ is determined only by the smallest eigenvalue $\lambda_{k}^{-}$
of the matrix $\mathbb{T}_{k}$,
\begin{equation}
\ln r^{(j)}=\frac{\alpha_{k}^{+}}{\zeta_{k}^{+}-\zeta_{k}^{-}}\left(\lambda_{k}^{-}\right)^{j}\qquad\mbox{and}\qquad\ln s^{(j)}=-\frac{\alpha_{k}^{+}\zeta_{k}^{-}}{\zeta_{k}^{+}-\zeta_{k}^{-}}\left(\lambda_{k}^{-}\right)^{j}.\label{eq:lnrscrit}
\end{equation}

Precisely at the critical point, there are exact real-space RG calculations
\cite{luck93a,igloi97b,hermisson97} that provide information on the
dynamic scaling between energy levels and length scales. These can
be used to gauge the reliability of the SDRG approach developed here.

\subsubsection{Irrelevant aperiodicity.}

For the Fibonacci chain, which corresponds to $k=1$, we have $\lambda_{k}^{-}\simeq0.382$,
so that both $r^{(j)}$ and $s^{(j)}$ asymptotically approach unity.
This means that the perturbative recursion relations become poorer
and poorer approximations as the iterations proceed, suggesting at
the same time, in agreement with known results (see \cite{luck93a}
and references therein), that the critical behavior should correspond
to that of the uniform chain. In other words, Fibonacci modulations
are irrelevant for the critical behavior of the quantum Ising chain,
and the relation between energy and length scales at criticality follows
the usual dynamical scaling form
\[
\Omega\sim\ell^{-z},
\]
with a dynamic exponent $z=1$.

\subsubsection{Marginal aperiodicity.}
\label{sub:marginal}

For $k=2$, we have $\lambda_{k}^{-}=1$, so that both $r^{(j)}$
and $s^{(j)}$ remain constant along the iterations, being expressed
in terms of the original bonds as
\[
r^{(j)}=\left(\frac{J_{a}}{J_{b}}\right)^{d_{a}}\qquad\mbox{and}\qquad s^{(j)}=\left(\frac{J_{a}}{J_{b}}\right)^{1-d_{a}},
\]
so that the ratio between the effective bonds is constant,
\[
\rho^{(j)}\equiv\frac{J_{a}^{(j)}}{J_{b}^{(j)}}=r^{(j)}s^{(j)}=\frac{J_{a}}{J_{b}}\equiv\rho.
\]
This corresponds to a line of critical points, associated with different
values of the ratio $\rho=J_{a}/J_{b}$, and gives rise to a nonuniversal
dynamic exponent $z$, as already known for other aperiodic sequences
\cite{luck93a,igloi97b,hermisson97}, and for quantum Ising chains
with correlated disorder \cite{hoyos2011}. Indeed, if we use the
values of the strong bonds at each step of the RG procedure to estimate
the gaps $\Omega_{j}$ between different energy levels, we get
\[
\Omega_{j}\sim J_{b}^{(j)}=J_{a}^{(j-1)}=\frac{J_{a}^{(j-1)}}{J_{b}^{(j-1)}}J_{b}^{(j-1)}=\rho J_{b}^{(j-1)}=\rho^{j}J_{b},
\]
while Eq. \ref{eq:lalblh} leads to a characteristic length
\[
\ell_{j}\sim\ell_{b}^{(j)}\sim\left(\lambda_{2}^{+}\right)^{j}=4^{j}.
\]
Expressing $j$ in terms of $\ell_{j}$, we then obtain the dynamic
scaling form
\begin{equation}
\Omega_{j}\sim\ell_{j}^{-z\left(\rho\right)},\label{eq:gapvslmarg}
\end{equation}
with a dynamic exponent
\begin{equation}
z\left(\rho\right)=-\frac{\ln\rho}{\ln4}.\label{eq:zkeq2sdrg}
\end{equation}

As in the case of the quantum spin-$\frac{1}{2}$ \emph{XX} chain
with marginal aperiodic bonds \cite{vieira05a,vieira05b}, the dynamic
exponent coming out of the SDRG approach corresponds to the leading
term in the $\rho\ll1$ expansion of an exact result which can be
obtained by other methods \cite{igloi97b,hermisson97,hermisson00}.
In the present case, the exact result turns out to be 
\begin{equation}
z_{\mathrm{ex}}\left(\rho\right)=\frac{\ln\left(\rho^{1/2}+\rho^{-1/2}\right)}{2},\label{eq:zkeq2exact}
\end{equation}
which can be obtained from the corresponding result for the quantum
Ising chain with bonds chosen from the period-doubling sequence generated
by the substitution rule
\[
\left\{ \begin{array}{l}
A\rightarrow AB\\
B\rightarrow AA
\end{array}\right..
\]
Distributing couplings $J_{A}$ and $J_{B}\gg J_{A}$ according to
the above sequence, and applying an intermediate field $h$, the SDRG
approach leads to a sequence of bonds $J_{a}$ and $J_{b}$ given
by the rule in Eq. \ref{eq:substitutionrule} with $k=2$. At criticality,
these bonds are related to the original ones via
\[
J_{a}=\frac{J_{A}^{5/3}}{J_{B}^{2/3}},\qquad J_{b}=J_{A},
\]
so that
\[
\frac{J_{a}}{J_{b}}=\left(\frac{J_{A}}{J_{B}}\right)^{2/3}.
\]
Thus, the exact result for the period-doubling chain \cite{igloi97b,hermisson97}
translates into Eq. \ref{eq:zkeq2exact}, which, for $\rho\ll1$,
clearly reduces to Eq. \ref{eq:zkeq2sdrg}.

\subsubsection{Relevant aperiodicity.}

For $k\geq3$, the wandering exponent $\omega_{k}$ is positive. According
to the Harris-Luck criterion, this corresponds to relevant aperiodicity. 

It is easy to check that, under the condition $k\geq3$, $\lambda_{k}^{-}$
is larger than unity. Since we start with $r^{(0)}<1$ and $s^{(0)}<1$,
we have $\alpha_{k}^{+}<0$, and Eq. \ref{eq:lnrscrit} implies that
both $r^{(j)}$ and $s^{(j)}$ become arbitrarily close to zero as
the RG scheme proceeds. This means that the perturbative recursion relations
become asymptotically exact, as does the RG treatment itself. This
is analogous to what happens for the random quantum Ising chain \cite{fisher92,fisher95}.
It is then not surprising that the activated dynamic scaling form
\begin{equation}
\Omega_{j}\sim\exp\left[-\left(\ell_{j}/\ell_{\rho}\right)^{\omega}\right],\label{eq:dynrelrelevant}
\end{equation}
with $\ell_{\rho}\sim\left|\ln\rho\right|^{-1/\omega}$ a length scale that depends on the original couplings,
is exactly reproduced by the SDRG treatment, as it can be readily
checked from Eqs. \ref{eq:jajbeff}, \ref{eq:hmueff}, \ref{eq:hck},
and \ref{eq:lalblh}. Thus, the dynamic exponent is formally infinite.
This result is analogous to what is verified for the critical spin-$\frac{1}{2}$
\emph{XXZ} chain with relevant aperiodic bonds \cite{vieira05a,vieira05b}.

Notice that $\omega$ here plays the role of the {}``tunneling''
exponent $\psi$ in the random quantum Ising chain \cite{fisher99},
which owes its name to the need of high-energy virtual excitations
in order to flip spin clusters.

\subsection{The neighborhood of the critical point}

In order to study the system in the neighborhood of the critical point,
we set 
\begin{equation}
h=h_{\mathrm{crit}}\left(1-\delta\right),\qquad\left|\delta\right|\ll1,
\label{eq:deltaaperiod}
\end{equation}
with $\delta$ measuring the distance to criticality. Taking into account
Eq. \ref{eq:hck}, and the fact that now $\overline{\ln J_{i}}=\ln h_{\mathrm{crit}}$,
one can see that the definitions of $\delta$ given by Eqs. \ref{eq:deltaaperiod} and
\ref{eq:deltarandom} are equivalent for $h\simeq h_{\mathrm{crit}}$.

It is clear
from the previous discussion that, for $\delta\neq0$, the RG process
has to be interrupted whenever the condition
\begin{equation}
\max\left\{ \ln r^{(j)},\ln s^{(j)}\right\} =0\label{eq:maxlnrlns}
\end{equation}
is reached, which happens for a value of $j$ we denote by $j_{\delta}$.
By using Eqs. \ref{eq:lnr0} and \ref{eq:lns0}, together with the
initial values 
\[
r^{(0)}=\frac{h}{J_{b}}=\rho^{d_{a}}\left(1-\delta\right)\qquad\mbox{and}\qquad s^{(0)}=\frac{J_{a}}{h}=\rho^{d_{b}}\left(1-\delta\right)^{-1},
\]
in which we again make use of the ratio $\rho=J_{a}/J_{b}$ between
the original bond values, we can show that for $\left|\delta\right|\ll1$
the condition in Eq. \ref{eq:maxlnrlns} leads to
\begin{equation}
j_{\delta}=\frac{\ln\left|\delta\right|}{\ln\left(\lambda_{k}^{-}/\lambda_{k}^{+}\right)}+c\rho\sim\frac{\ln\left|\delta\right|}{\ln\left(\lambda_{k}^{-}/\lambda_{k}^{+}\right)},\label{eq:jstarvseps}
\end{equation}
$c$ being a constant which is different in the paramagnetic ($\delta<0$)
and in the ferromagnetic ($\delta>0$) phases.

This last result allows us to obtain analytical expressions for critical
exponents, as discussed below. Of course, the reported results are
not valid for $k=1$, where the critical behavior corresponds to that
of the uniform chain.

\subsubsection{The paramagnetic phase.}

The paramagnetic phase is characterized by the fact that after $j_{\delta}$
iterations all effective bonds are smaller than the effective fields.
Thus, the length of the largest effective bonds at this point of the
RG process provides an estimate of the correlation length. 

Substituting Eq. \ref{eq:jstarvseps} into Eq. \ref{eq:lalblh}, we
obtain
\[
\xi\sim\ell_{b}^{\left(j_{\delta}\right)}\sim\left(\lambda_{k}^{+}\right)^{j_{\delta}}\sim\left|\delta\right|^{-\nu},
\]
with a correlation-length exponent
\begin{equation}
\nu=\frac{\ln\lambda_{k}^{+}}{\ln\lambda_{k}^{+}-\ln\lambda_{k}^{-}}.\label{eq:nuk}
\end{equation}
As $\lambda_{k}^{\pm}=\left(\zeta_{k}^{\pm}\right)^{2}$, this last
result can be rewritten as
\begin{equation}
\nu=\frac{1}{1-\omega_{k}}.\label{eq:nuk-1}
\end{equation}
We thus see that the correlation-length exponent at the aperiodic
critical point saturates the Harris-Luck criterion in the form
\begin{equation}
\nu\geq\frac{1}{1-\omega},\label{eq:invertHL}
\end{equation}
for all sequences in the family generated by the substitution rule
in Eq. \ref{eq:substitutionrule}. In particular, for the marginal
case of $k=2$ we recover the exact result $\nu=1$ \cite{igloi97b,hermisson97}.
We note that Eq. \ref{eq:invertHL} is analogous to the Chayes bound
for the correlation length critical exponent in the presence of disorder
\cite{chayes86}.

We can also study the behavior of the energy gap as a function of
$\delta$. From the discussion in Subsection \ref{sub:critical-point},
we see that the energy gap vanishes at the critical point, since $\Omega_{j}\rightarrow0$
as $j\rightarrow\infty$; see Eqs. \ref{eq:gapvslmarg} and \ref{eq:dynrelrelevant}.
On the other hand, in the paramagnetic phase the value of the effective
field $h^{\left(j\right)}$ for $j=j_{\delta}$, being much larger
than any other effective coupling, yields an estimate of the energy
gap between the ground state and the first excited state of the infinite
chain. We thus have
\[
\Omega\left(\delta\right)\sim h^{\left(j_{\delta}\right)}\sim\left|\delta\right|^{z\left(\rho\right)\nu},
\]
for marginal aperiodicity ($k=2$), and,
\begin{equation}
\Omega\left(\delta\right)\sim h^{\left(j_{\delta}\right)}\sim\exp\left(-C\left|\delta\right|^{-\omega\nu}\right),\label{eq:gapvseps}
\end{equation}
for relevant aperiodicity ($C$ being a nonuniversal constant). The
above expressions can be derived from the recursion relations in Eqs.
\ref{eq:jajbeff} and \ref{eq:hmueff}, with the help of Eq. \ref{eq:jstarvseps}.
Notice that $\Omega\left(\delta\right)\rightarrow0$ as $\delta\rightarrow0$,
but the noncritical gap is always finite, implying the absence of
Griffiths singularities in the presence of aperiodic fluctuations
generated by substitution rules \cite{igloi98b,hermisson00}.

\subsubsection{The ferromagnetic phase.}

In the ferromagnetic phase, after $j_{\delta}$ iterations of the
RG transformation the effective field becomes smaller than both effective
bonds. An estimate of the average spontaneous magnetization is provided
by the average magnetic moment of all the spins still active at this
point of the RG procedure (\emph{i.e.} those that have not been decimated
under the action of strong local fields in the previous RG steps).
Since the sequence of bonds remains invariant up to this point, the
estimate is simply given by
\[
m_{x}\sim n_{j_{\delta}}\mu^{\left(j_{\delta}\right)}\sim\frac{\mu^{\left(j_{\delta}\right)}}{\ell_{j_{\delta}}},
\]
in which $n_{j}$ is the fraction of active spins after $j$ iterations
of the RG transformation. From Eq. \ref{eq:hmueff}, we see that
\[
\mu^{(j)}=\left(k+1\right)^{j}\mu_{0},
\]
yielding
\[
m_{x}\sim\left(\frac{k+1}{\lambda_{k}^{+}}\right)^{j_{\delta}}.
\]
Substituting Eq. \ref{eq:jstarvseps} then leads to
\[
m_{x}\sim\delta^{\beta},
\]
with a critical exponent
\begin{equation}
\beta=\frac{\ln\left(k+1\right)-\ln\lambda_{k}^{+}}{\ln\lambda_{k}^{-}-\ln\lambda_{k}^{+}}.\label{eq:betak}
\end{equation}

Notice that, if we follow Fisher \cite{fisher92,fisher95} and define
at the critical point an exponent $\phi$ relating the effective magnetic
moment to the energy scale,
\[
\mu^{\left(j\right)}\sim\left[\ln\frac{\Omega_{\rho}}{\Omega_{j}}\right]^{\phi},
\]
where $\Omega_{\rho}$ is a nonuniversal constant, we find	 
\[
\beta=\nu\left(1-\phi\psi\right),
\]
just as for the random quantum Ising chain \cite{fisher99}.

\subsection{Critical correlation functions}

\label{sub:Corrfunc}Now we determine the asymptotic behavior of the
pair correlation function associated with the order parameter,
\[
C^{xx}\left(l\right)=\overline{\left\langle \sigma_{i}^{x}\sigma_{i+l}^{x}\right\rangle },
\]
in which the bar denotes average over all lattice sites $i$. At criticality,
this correlation function follows a power-law scaling form,
\[
C^{xx}\left(l\right)\sim\frac{1}{l^{\eta}}.
\]
In the uniform limit, and thus also for the Fibonacci chain, the exponent
$\eta$ takes the value $\frac{1}{4}$ \cite{barouch71}. Below we
calculate $\eta$ for $k\geq2$.

The elements associated with the strongest bonds $J_{b}^{\left(j\right)}$
in each RG step are clusters containing $k+1$ spins. Each of these
clusters, if decoupled from the rest of the chain, has a doubly degenerate
ground state (in the absence of a field), corresponding to all spins
pointing along the $+x$ or the $-x$ direction. If we denote these
states by $\left|\Psi^{\pm}\right\rangle $, it is clear that
\[
\left\langle \Psi^{\pm}\left|\sigma_{i}^{x}\right|\Psi^{\pm}\right\rangle =\pm1,
\]
in which $i$ labels the spins inside a given cluster.

\begin{figure}
\begin{centering}
\includegraphics[angle=90,height=0.9\textheight]{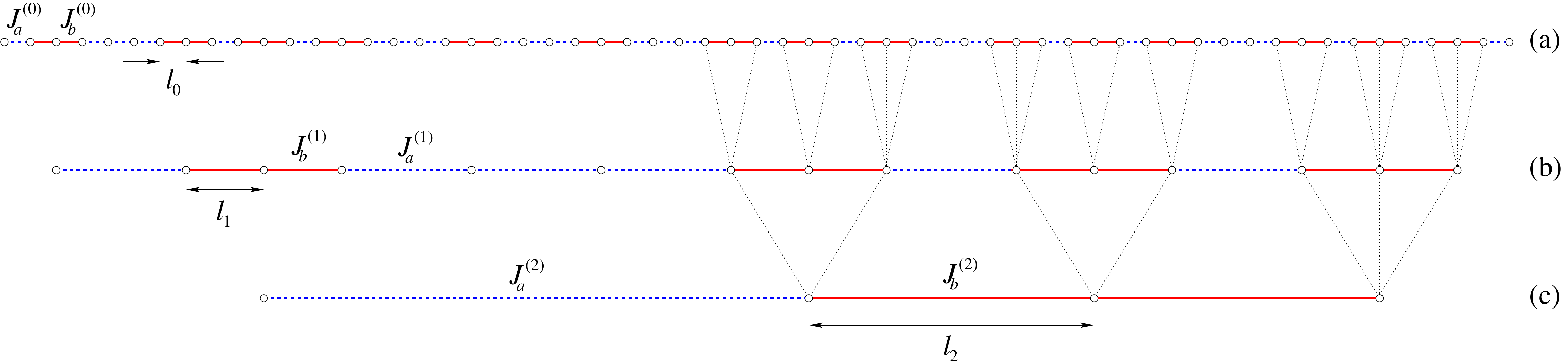}
\par\end{centering}

\caption{\label{fig:corrfunck}Estimating correlation functions for the case
$k=2$.}

\end{figure}
Owing to the inflation symmetry associated with couplings generated
by the substitution rule in Eq. \ref{eq:substitutionrule}, these
clusters are either grouped into superclusters of $k+1$ clusters
or are separated from each other by $k+1$ weak bonds $J_{a}$, as illustrated
for $k=2$ in Fig. \ref{fig:corrfunck}(a). The superclusters, on
the other hand, are either grouped into ``supersuperclusters''
containing $k+1$ superclusters or isolated from each other by clusters
and weak bonds, and so on. Thus, there appears a hierarchy of $\left(k+1\right)$-spin
clusters, yielding a self-similar structure of the correlation function,
and leading to the appearance of characteristic distances where the
correlation is particularly strong.

In order to analyze this structure, we recourse to the simplest approximation,
in which we assume that spins appearing in the same cluster at some
step of the RG scheme are maximally correlated, while the remaining
correlations are negligible. Looking at Fig. \ref{fig:corrfunck}(a),
we see that, in the limit of a very large chain (where boundary effects
can be ignored), the average correlation between spins separated by
the distance $l_{0}\equiv\ell_{b}^{(0)}+\ell_{h}^{(0)}=1$ , corresponding
to the real distance between nearest-neighbor spins connected by a
coupling $J_{b}$, is given by
\[
C^{xx}\left(l_{0}\right)=\frac{1}{N}\times k\times\frac{N_{b}^{(0)}}{k}=\frac{N_{b}^{(0)}}{N},
\]
in which $N$ is the total number of spins, $N_{b}^{(0)}$ is the
number of $J_{b}$ bonds ($\frac{1}{k}N_{b}^{(0)}$ being the number
of clusters with $k+1$ spins), and the multiplicative $k$ factor
comes from the fact that in each cluster there are $k$ pairs of spins
separated by the distance $l_{0}$. In fact, we have
\[
C^{xx}\left(l_{0}^{\prime}\right)=\frac{k+1-g_{0}}{k}\times\frac{N_{b}^{(0)}}{N},
\]
since in each cluster there are $k+1-g_{0}$ pairs of spins separated
by the distance $l_{0}^{\prime}=g_{0}l_{0}$, $g_{0}\in\left\{ 1,2,\dots,k\right\} $.
However, the strongest correlation corresponds to the value $g_{0}=1$,
coming from spins separated by the distance $l_{0}^{\prime}=l_{0}$.

From Fig. \ref{fig:corrfunck}(b), we see that an estimate of the
average correlation between spins separated the distance $l_{1}\equiv\ell_{b}^{(1)}+\ell_{h}^{(1)}=k+1$
is given by
\[
C^{xx}\left(l_{1}\right)=\left[\frac{1}{N}\times k\times\frac{N_{b}^{(1)}}{k}\right]\times\left(k+1\right)=\left(k+1\right)\frac{N_{b}^{(1)}}{N},
\]
with $N_{b}^{(1)}$ denoting the number of effective bonds $J_{b}^{(1)}$
and the multiplicative factor $k+1$ coming from the fact that each
effective spin represents $k+1$ real spins in this step of the hierarchy.
Notice that inside each cluster there are in fact contributions to
the average correlation between real spins separated by distances
$l_{1}^{\prime}=g_{1}l_{1}\pm g_{0}l_{0}$, where $g_{1}\in\left\{ 1,2,\dots,k\right\} $
and $g_{0}\in\left\{ 0,1,\dots,k\right\} $. (However, these correlations
are strongest for $l_{1}^{\prime}=l_{1}$.) Since $l_{1}-kl_{0}=1$,
$l_{1}-\left(k-1\right)l_{0}=2$, $\dots$, $l_{1}-l_{0}=k$, this
gives corrections to the average correlations $C^{xx}\left(l_{0}\right)$,
$C^{xx}\left(2l_{0}\right)$, $\dots$, $C^{xx}\left(kl_{0}\right)$
between spins separated by short distances in the real chain. 

Figure \ref{fig:corrfunck}(c) leads us to conclude that the average
correlation between real spins separated by the distance $l_{2}\equiv\ell_{b}^{(2)}+\ell_{h}^{(2)}=k^{2}+3k+1$
is approximately given by
\[
C^{xx}\left(l_{1}\right)=\left[\frac{1}{N}\times k\times\frac{N_{b}^{(2)}}{k}\right]\times\left(k+1\right)^{2}=\left(k+1\right)^{2}\frac{N_{b}^{(2)}}{N},
\]
since now every effective spin represents $\left(k+1\right)^{2}$
real spins. In fact, there are contributions to the average correlations
between real spins separated by distances $l_{2}^{\prime}=g_{2}l_{2}\pm g_{1}l_{1}\pm g_{0}l_{0}$,
in which $g_{2}\in\left\{ 1,2,\dots,k\right\} $ and $g_{1},g_{0}\in\left\{ 0,1,\dots,k\right\} $.
(Again, the strongest contribution comes from spins separated by the
distance $l_{2}^{\prime}=l_{2}$.) Here there are corrections to correlations
between spins separated by distances of the order of $l_{1}$, but
not of the order of $l_{0}$, since $l_{2}-kl_{1}-kl_{0}=l_{1}$ corresponds
to the minimum value of $l_{2}^{\prime}$.

Generalizing the above discussion, there are strong correlations 
associated with distances
around $l_{j}\equiv\ell_{b}^{(j)}+\ell_{h}^{(j)}\sim\left(\lambda_{k}^{+}\right)^{j}$,
with
\[
C^{xx}\left(l_{j}\right)=\left(k+1\right)^{j}\frac{N_{b}^{(j)}}{N},
\]
$N_{b}^{\left(j\right)}$ being the number of effective bonds $J_{b}^{\left(j\right)}$
in a chain with $N\gg1$ sites. Actually, there are contributions
to correlations between real spins separated by distances 
\[
l_{j}^{\prime}=g_{j}l_{j}\pm g_{j-1}l_{j-1}\pm\cdots\pm g_{0}l_{0},
\]
with $g_{j}\in\left\{ 1,2,\dots,k\right\} $ and $g_{i\neq j}\in\left\{ 0,1,\dots,k\right\} $,
so that the strongest contribution appears for $l_{j}^{\prime}=l_{j}$,
and the minimum value of $l_{j}^{\prime}$ is of the order of $l_{j-1}$.
Thus, there are no corrections for correlations between spins separated
by distances of the order of $l_{j-2}$ and below. 

Owing to the invariance of the sequence of effective bonds along the
RG steps, the number of effective $J_{b}^{\left(j\right)}$ bonds
scales as
\[
N_{b}^{\left(j\right)}\sim\frac{N}{l_{j}}.
\]
It is then clear that we can estimate the long-distance scaling behavior
of the average strong correlations as
\begin{equation}
C^{xx}\left(l_{j}\right)\sim\frac{\left(k+1\right)^{j}}{l_{j}}\sim\frac{1}{l_{j}^{\eta}},
\label{eq:corrxx}
\end{equation}
with an exponent
\begin{equation}
\eta=1-\frac{\ln\left(k+1\right)}{\ln\lambda_{k}^{+}}=\frac{\beta}{\nu}.\label{eq:etak}
\end{equation}
This result is in contrast with the one obtained for the random quantum
Ising chain, for which critical correlations decay with distance as
$C^{xx}\left(l\right)\sim l^{-2\beta/\nu}$ \cite{fisher95,igloi2002},
due to the probabilistic nature of cluster formation.

We note that the prediction in Eq. \ref{eq:corrxx} is valid only for the characteristic
distances $l_j$. Correlations between spins separated by generic distances (distinct from the 
$l_{j}^{\prime}$ defined above) are expected to be much weaker. In fact, as shown explicitly
for the \emph{XX} chain in Ref. \cite{vieira05b} (Appendix B), 
these weaker correlations are proportional to
the effective couplings, and thus follow essentially the same scaling form as 
in Eq. \ref{eq:gapvslmarg} (\ref{eq:dynrelrelevant}) for marginal (relevant) aperiodicity.

\subsection{Numerical calculations}

The analytical results derived in the previous subsections can be
checked by numerical calculations based on the well-known mapping
of the quantum Ising chain onto a free-fermion system \cite{katsura62,pfeuty70}.
Here we use the recipe described in Ref. \cite{igloi98c}. The numerical
work then involves diagonalizing matrices of order $2L$ for chains
containing $L$ spins. As here we work with open boundary conditions,
this corresponds to $L-1$ bonds.

When dealing with finite chains, we have to take into account that
substitution rules produce aperiodic systems containing many different
subsequences of the same length $L-1$. We then have to perform a
weighted average over all $O\left(L\right)$ such subsequences, with
weights given by the frequencies with which a given subsequence appears
in the infinite sequence. For arbitrary $L$, in order to determine
these frequencies we iterate the substitution rule until a very large
sequence of size $\mathcal{L}$ (with $\mathcal{L}$ ranging from
$\sim10^{4}$ to $\sim10^{6}$) is obtained. We then calculate the
finite-size estimates of the corresponding frequencies, and extrapolate
to infinite $\mathcal{L}$ using standard algorithms \cite{guttmann89}.
In order to minimize fluctuations due to the discrete scale invariance
exhibited by substitutional sequences, the number of bonds, $L-1$,
is always chosen to coincide with the length of the aperiodic sequences
after an integer number of iterations of the substitution rule.

We work here in the so-called canonical ensemble of disorder, according
to which we study the critical properties of the system by performing
numerical calculations for a value of the transverse field corresponding
to the critical field of the infinite sequence. The other possibility,
which defines the microcanonical ensemble, would correspond to setting
the transverse field to values calculated from the densities of letters
$a$ and $b$ in each particular subsequence of finite length, which
are usually distinct from the infinite-sequence value. It has been
shown that the canonical ensemble yields the same critical exponents
as exact and RG calculations \cite{igloi98c}.

The finite-size behavior of the bulk critical magnetization is described
by
\[
m_{x}\left(L\right)\sim L^{-\beta/\nu},
\]
for systems of size $L$. Within the free-fermion approach, this scaling
behavior can be checked by calculating magnetization profiles (i.e.
local expectation values of the spin operators $\sigma_{l}^{x}$,
$l=1,2,\ldots,L$) in finite chains with prescribed boundary conditions
\cite{igloi98c}. Here we focus on fixed boundary conditions, corresponding
to $h_{1}=0$ and $h_{L}=0$, and look at the resulting profiles when
both end spins are in the state $\left|\sigma_{i}^{x}\right\rangle =\left|+\right\rangle $.

\begin{figure}
\begin{centering}
\includegraphics[width=0.75\textwidth]{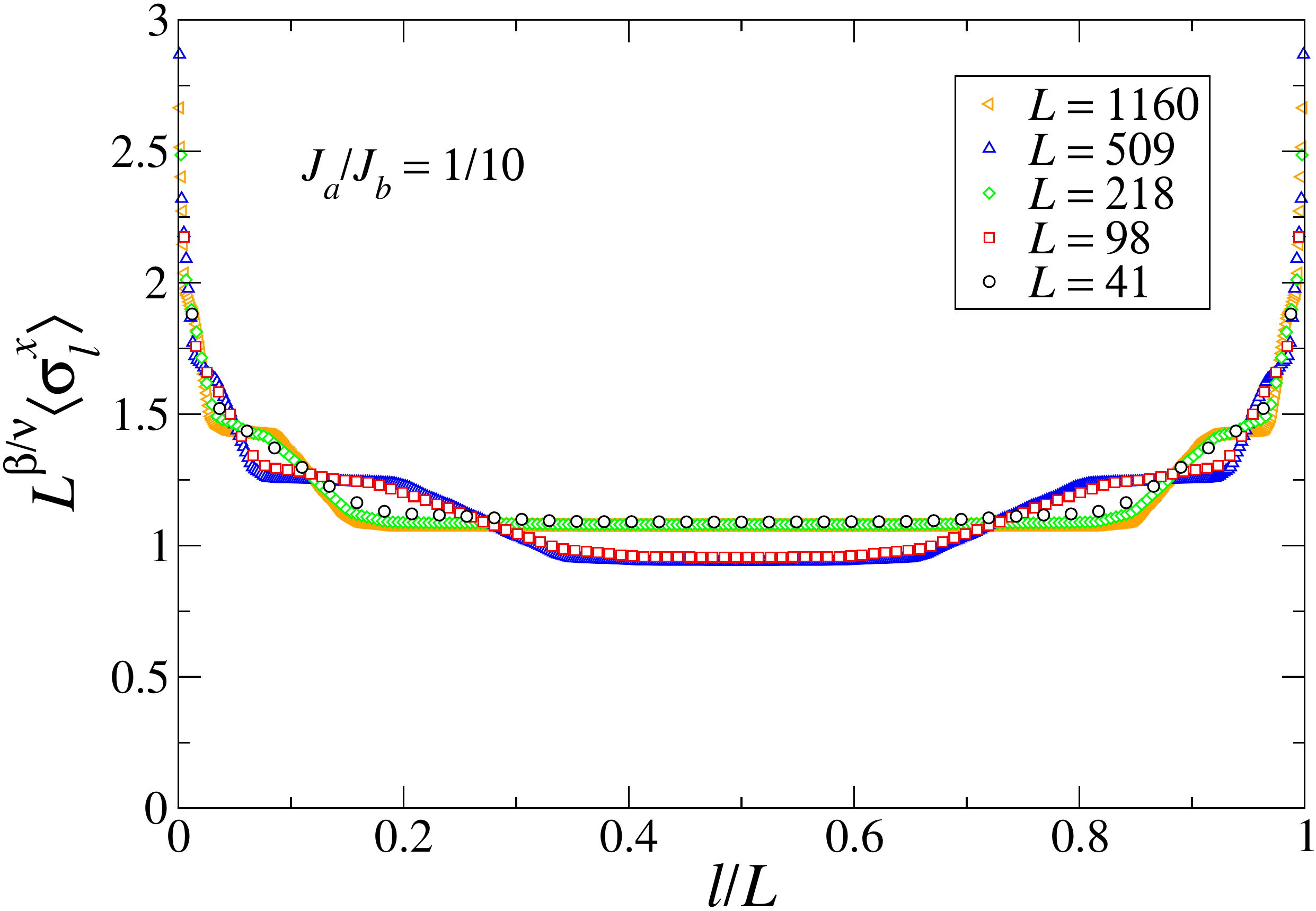}
\par\end{centering}

\caption{\label{fig:magprofk3}Rescaled magnetization profiles for $k=3$ at
the critical point, for different chain lengths $L$. The values of
the exponents $\beta$ and $\nu$ were set to their SDRG predictions,
yielding $\beta/\nu\simeq 0.169$. Calculations were performed
for fixed ends, $\left\langle \sigma_{1}^{x}\right\rangle =\left\langle \sigma_{L}^{x}\right\rangle =+1$,
and the curves result from averaging over all distinct subsequences
of length $L$, followed by mirror symmetrization around the position
$l=L/2$.}
\end{figure}
For $k=3$, a case in which aperiodic modulations are relevant, we
expect the SDRG prediction for $\beta/\nu$ to be valid for any coupling
ratio $\rho=J{}_{a}/J_{b}$. Of course for $J_{a}/J_{b}$ close to
unity there will be a large crossover length $L_{\times}$, below
which the system will appear uniform, and the SDRG result should be
recovered only for $L>L_{\times}$, while for smaller values of $\rho$
the SDRG should be evident even for relatively small chain lengths.
This is confirmed by the free-fermion calculations, as shown in Fig.
\ref{fig:magprofk3} for $J_{a}/J_{b}=1/10$. Notice that the results
for finite $L$ ($L$ between $41$ and $1160$) collapse onto two
different master curves, depending on whether the subsequence size
was produced after an odd or an even number of iterations of the substitution
rule. This is related to the fact that the deviation of the number
of letters $a$ (or equivalently $b$) with respect to the expectation
value in the infinite sequence oscillates in sign between consecutive
iterations of the substitution rule (the eigenvalue $\zeta_{2}$ of
the substitution matrix is negative). Furthermore, it should be noted
that none of the master curves follows the conformal invariance prediction
\cite{burkhardt85,igloi93}
\[
m_{x}\left(l/L\right)=\left\langle \sigma_{l}^{x}\right\rangle \sim\left[L\sin\left(\pi l/L\right)\right]^{-\beta/\nu},
\]
as expected from the strongly anisotropic scaling along the space
and time directions implied by Eq. \ref{eq:dynrelrelevant}. 

\begin{figure}
\begin{centering}
\includegraphics[width=0.75\textwidth]{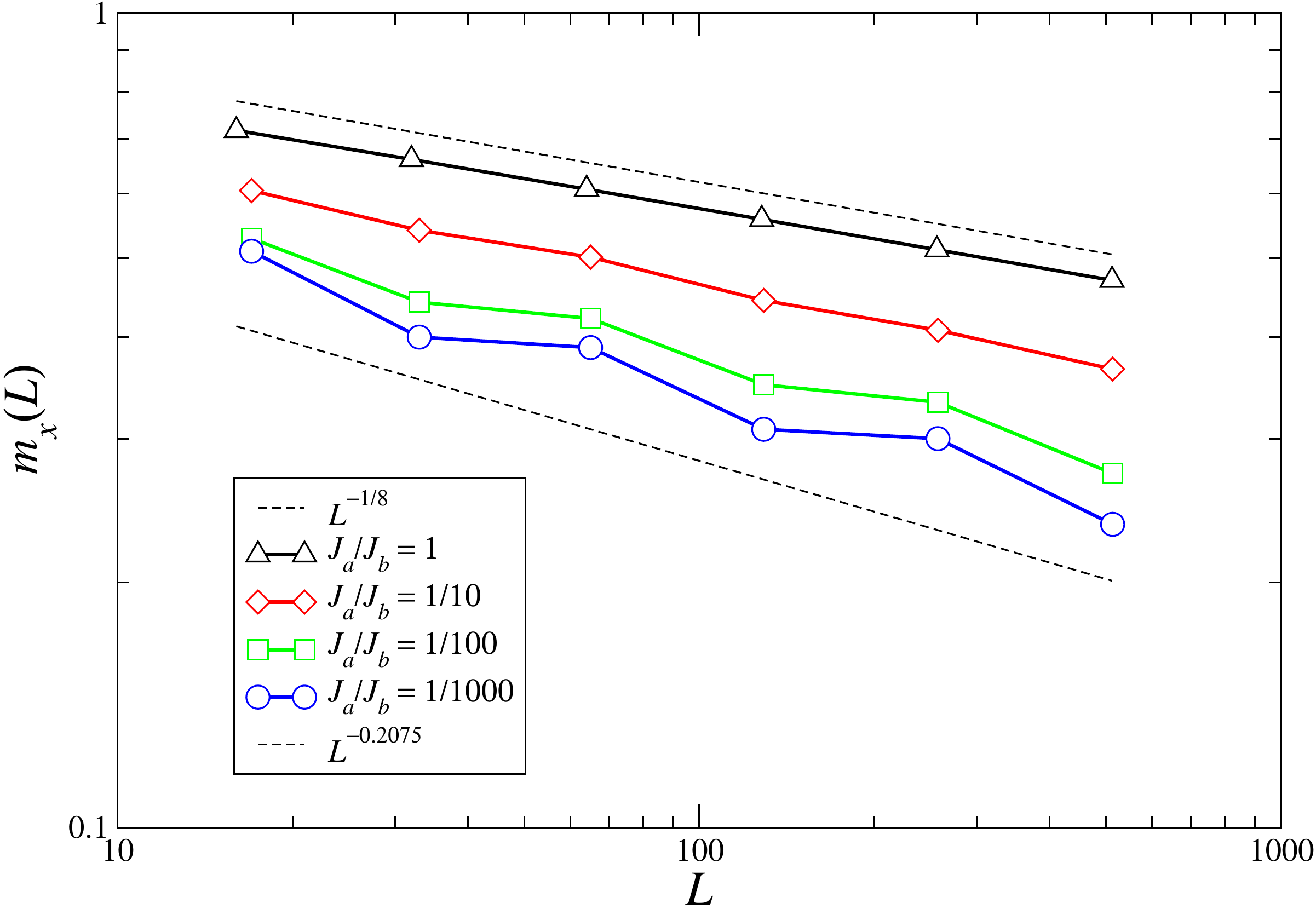}
\par\end{centering}

\caption{\label{fig:magvsLk2}Critical bulk magnetization for $k=2$
(actually, for the related period-doubling chain described in
\ref{sub:marginal}) as a function
of the chain length $L$. The upper dashed curves correspond to the
Onsager behavior $L^{-\beta/\nu}$, with $\beta/\nu=1/8$, while the
lower dashed curve indicates the SDRG prediction, $\beta/\nu\simeq0.2075$. }

\end{figure}
In the marginal case, $k=2$, we only expect the SDRG predictions
to be satisfied for coupling ratios approaching zero. Indeed, as shown
in Fig. \ref{fig:magvsLk2}, a plot of the finite-size behavior of
the bulk magnetization at criticality (estimated from the corresponding
value in the middle of the chain) shows a clear change from the Onsager
universality class to the strong-modulation behavior as the coupling
ratio is reduced. 

\begin{figure}
\begin{centering}
\includegraphics[width=0.75\textwidth]{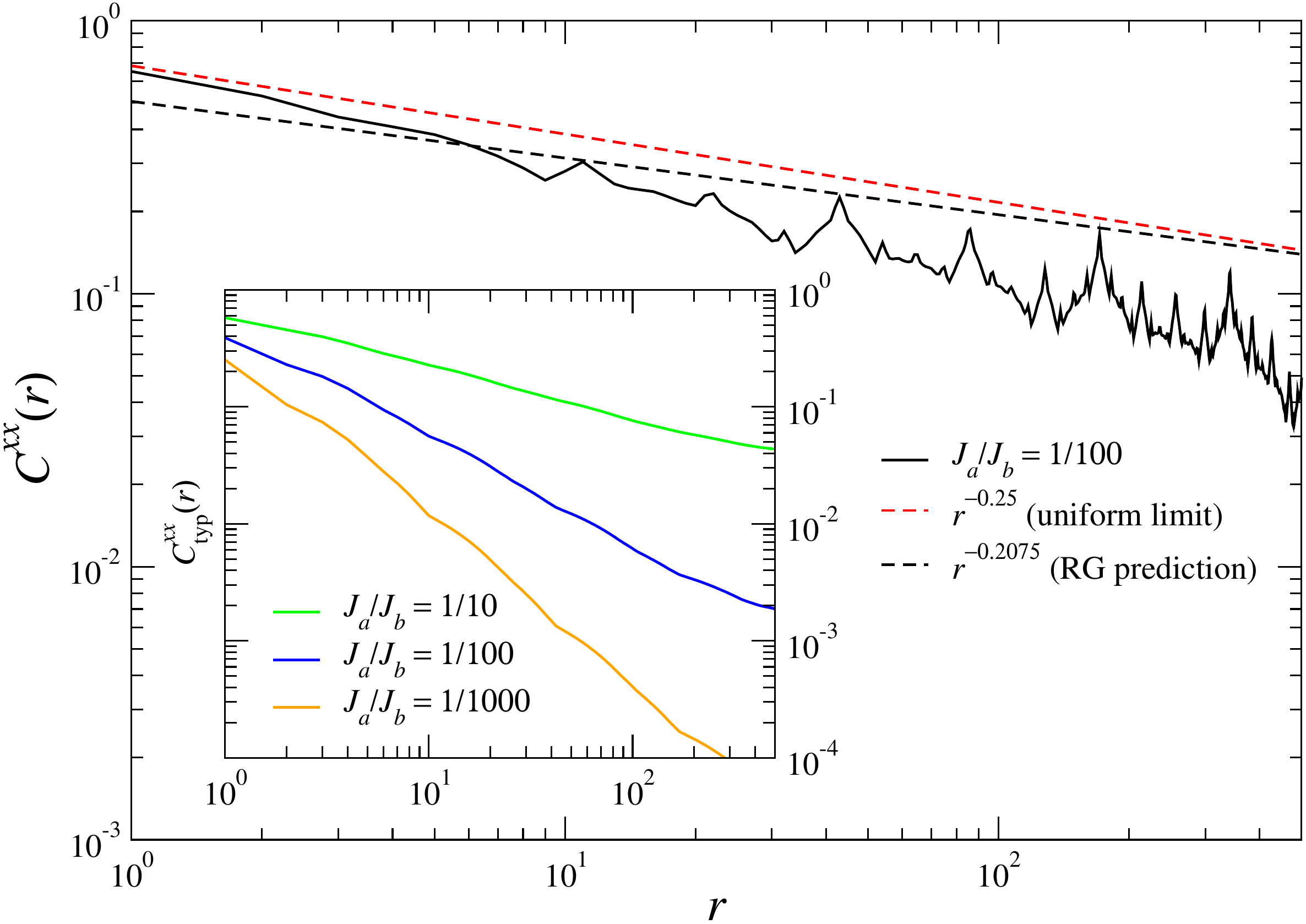}
\par\end{centering}

\caption{\label{fig:corrk2}Main panel: Average pair correlation function versus
spin separation for $k=2$. The red dashed curve shows the Onsager
behavior, while the black dashed curve indicates the SDRG prediction,
which is valid at the characteristic distances where correlations
are strong. Inset: Typical correlation function for various values
of the coupling ratio. In all cases, numerical calculations correspond
to chains with $1366$ sites.}

\end{figure}
The behavior of the critical pair correlation function $C^{xx}\left(l\right)$,
as obtained from the free-fermion method, is illustrated in the main
panel of Fig. \ref{fig:corrk2} for $k=2$ and a small coupling ratio
$J_{a}/J_{b}=1/100$. Notice the existence of characteristic distances
for which spins are strongly correlated, due to the presence of strong
effective bonds, and of a ultrametric structure produced by the cluster
hierarchy discussed in Subsection \ref{sub:Corrfunc}. The inset shows
plots of the typical correlation, estimated by
\[
C_{\mathrm{typ}}^{xx}\left(l\right)\equiv\exp\left(\overline{\ln\left\langle \sigma_{i}^{x}\sigma_{i+l}^{x}\right\rangle }\right),
\]
which exhibits nonuniversal behavior and a much smoother dependence
on the spin separation $l$, confirming that, as in the random quantum
Ising chain, there is a decoupling between average and typical behavior. 
As contributions from the rare strongly correlated spin pairs 
are washed out by the logarithm in the
calculation of $C_{\mathrm{typ}}^{xx}$, typical correlations behave
just as the weak correlations discussed in the end of subsection
\ref{sub:Corrfunc}.

\section{\label{sec:Rudin-Shapiro-sequences}Rudin-Shapiro sequences}

\subsection{The four-letter sequence}

The family of sequences analyzed in the previous section has the property
that, at the criticality, there is only one type of spin cluster connected
by strong bonds at a given step of the RG process. A more complex
situation is produced by choosing bonds according to the four-letter
Rudin-Shapiro (RS) sequence, generated by the substitution rule 
\begin{equation}
\left\{ \begin{array}{c}
a\rightarrow ab\\
b\rightarrow ac\\
c\rightarrow db\\
d\rightarrow dc
\end{array}\right..\label{eq:rs4subst}
\end{equation}
The associated substitution matrix has eigenvalues $0$, $\pm\sqrt{2}$,
and $2$, so that the wandering exponent is $\omega=\frac{1}{2}$,
the same produced by uncorrelated randomness. In the resulting infinite
sequence all letters appear with the same frequency, and no clusters
of repeated letters are present. 

\begin{figure}
\begin{centering}
\includegraphics[angle=90,height=0.7\textheight]{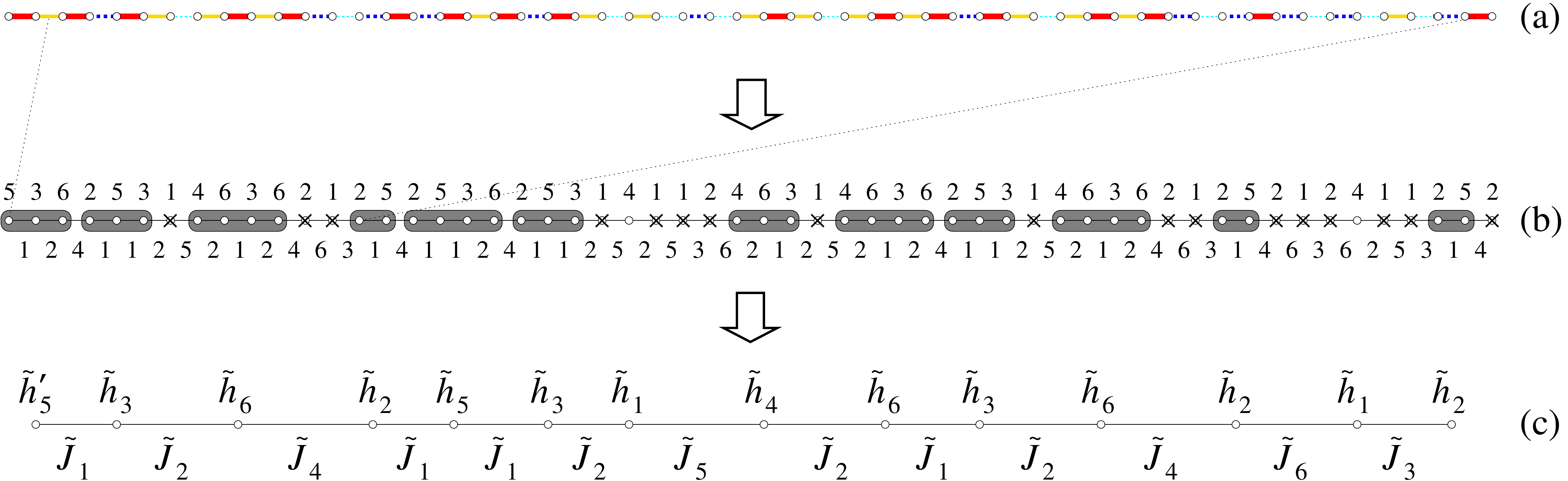}
\par\end{centering}

\caption{\label{fig:rs4sdrg}Renormalization-group transformation as applied
to the four-letter Rudin-Shapiro chain. The original chain is shown
in (a): $J_{a}$ bonds are represented by thick solid lines (red);
$J_{b}$, by solid lines (yellow); $J_{c}$, by thick dashed lines
(blue); $J_{d}$, by thin dashed lines (cyan); open circles mark the
position of the spins, under a uniform field $h\simeq h_{\mathrm{crit}}$.
Spins connected by $J_{a}$ or $J_{b}$ bonds form clusters, while
spins connected only by $J_{c}$ or $J_{d}$ are decimated. These
processes give rise to the effective chain in (b), characterized by
6 different values of both bonds and fields, respectively indicated
by the numbers below and above the chain. Upon further lowering of
the energy scale, effective spins under the grey boxes cluster together, 
while crossed circles denote effective spins decimated
by the action of strong effective fields. As shown in (c), apart from
a minor boundary effect, the next effective chain preserves the structure
in (b). }
\end{figure}
We choose couplings such that $J_{b}=\rho J_{a}$, $J_{c}=\rho^{2}J_{a},$and
$J_{d}=\rho^{3}J_{d}$, with $0<\rho<1$. Based on Pfeuty's result
\cite{pfeuty79}, we anticipate a critical field
\[
h_{\mathrm{crit}}=\rho^{\frac{3}{2}}J_{a},
\]
so that, in order to look at the system in the neighborhood of the
critical point, we focus on the condition $J_{d}\ll J_{c}\ll h\ll J_{b}\ll J_{a}$.
At energy scales $\Omega$ such that $J_{d}<J_{c}<\Omega<h<J_{b}<J_{a}$,
all spin pairs connected by $J_{a}$ or $J{}_{b}$ bonds behave as
single magnetic moments, whereas spins connected to their neighbors
at both sides by only $J_{c}$ or $J_{d}$ bonds are effectively removed
from the chain by the action of the transverse field; see Fig. \ref{fig:rs4sdrg}(a).
The resulting effective chain is composed of $6$ different bonds
and fields, whose values (in decreasing order of intensity), and those
of the corresponding effective magnetic moments, are given by
\begin{equation}
\begin{array}{lll}
J_{1} & = & J_{c},\\
J_{2} & = & J_{d},\\
J_{3} & = & J_{c}J_{d}/h,\\
J_{4} & = & J_{c}^{2}J_{d}/h^{2},\\
J_{5} & = & J_{c}J_{d}^{2}/h^{2},\\
J_{6} & = & J_{c}^{2}J_{d}^{2}/h^{3},
\end{array}\ \begin{array}{lll}
h_{1} & = & h^{2}/J_{b},\\
h_{2} & = & h^{2}/J_{a},\\
h_{3} & = & h^{3}/J_{a}J_{b},\\
h_{4} & = & h^{4}/J_{a}J_{b}^{2},\\
h_{5} & = & h^{4}/J_{a}^{2}J_{b},\\
h_{6} & = & h^{5}/J_{a}^{2}J_{b}^{2},
\end{array}\ \begin{array}{lll}
\mu_{1} & = & 2,\\
\mu_{2} & = & 2,\\
\mu_{3} & = & 3,\\
\mu_{4} & = & 4,\\
\mu_{5} & = & 4,\\
\mu_{6} & = & 5.
\end{array}\label{eq:jnhnrs4}
\end{equation}
As illustrated in Fig. \ref{fig:rs4sdrg}(b), weak bonds appear next
to strong fields and vice-versa. 

Writing again
\begin{equation}
h=h_{\mathrm{crit}}\left(1-\delta\right),\label{eq:hcrs4}
\end{equation}
which defines $\delta$ as a measure of the distance to criticality,
it can be seen that for $\left|\delta\right|\ll1$ and $\rho\ll1$
the couplings are ordered according to
\[
J_{1}\simeq h_{1}\gg J_{2}\simeq h_{2}\gg J_{3}\simeq h_{3}\gg J_{4}\simeq h_{4}\gg J_{5}\simeq h_{5}\gg J_{6}\simeq h_{6}.
\]
Sufficiently close to the critical point, upon further reduction of
the energy scale, this sequence of bonds and fields is mapped onto
itself by the bond recursion relations
\begin{equation}
\begin{array}{lll}
J_{1}^{\left(j+1\right)} & = & J_{4}^{\left(j\right)},\\
J_{2}^{\left(j+1\right)} & = & J_{2}^{\left(j\right)}J_{5}^{\left(j\right)}/h_{1}^{\left(j\right)},\\
J_{3}^{\left(j+1\right)} & = & J_{2}^{\left(j\right)}J_{3}^{\left(j\right)}J_{5}^{\left(j\right)}/\left[h_{1}^{\left(j\right)}\right]^{2},\\
J_{4}^{\left(j+1\right)} & = & J_{3}^{\left(j\right)}J_{4}^{\left(j\right)}J_{6}^{\left(j\right)}/h_{1}^{\left(j\right)}h_{2}^{\left(j\right)},\\
J_{5}^{\left(j+1\right)} & = & J_{2}^{\left(j\right)}J_{3}^{\left(j\right)}J_{5}^{\left(j\right)}J_{6}^{\left(j\right)}/\left[h_{1}^{\left(j\right)}\right]^{2}h_{2}^{\left(j\right)},\\
J_{6}^{\left(j+1\right)} & = & J_{3}^{\left(j\right)}J_{4}^{\left(j\right)}\left[J_{6}^{\left(j\right)}\right]^{2}/h_{1}^{\left(j\right)}\left[h{}_{2}^{\left(j\right)}\right]^{2},
\end{array}\label{eq:jeffrs4}
\end{equation}
the field recursion relations corresponding to the dual versions ($h_{i}\leftrightarrow J_{i}$)
of the above equations, thus defining the RG transformation. We also
write recursion relations for the effective magnetic moments, 
\begin{equation}
\begin{array}{lll}
\mu_{1}^{\left(j+1\right)} & = & \mu_{4}^{\left(j\right)},\\
\mu_{2}^{\left(j+1\right)} & = & \mu_{2}^{\left(j\right)}+\mu_{5}^{\left(j\right)},\\
\mu_{3}^{\left(j+1\right)} & = & \mu_{2}^{\left(j\right)}+\mu_{3}^{\left(j\right)}+\mu_{5}^{\left(j\right)},\\
\mu_{4}^{\left(j+1\right)} & = & \mu_{3}^{\left(j\right)}+\mu_{4}^{\left(j\right)}+\mu_{6}^{\left(j\right)},\\
\mu_{5}^{\left(j+1\right)} & = & \mu_{2}^{\left(j\right)}+\mu_{3}^{\left(j\right)}+\mu_{5}^{\left(j\right)}+\mu_{6}^{\left(j\right)},\\
\mu_{6}^{\left(j+1\right)} & = & \mu_{3}^{\left(j\right)}+\mu_{4}^{\left(j\right)}+2\mu_{6}^{\left(j\right)},
\end{array}\label{eq:mueffrs4}
\end{equation}
as well as for the lengths of the effective bonds and fields, 
\begin{equation}
\begin{array}{lll}
\ell_{J1}^{\left(j+1\right)} & = & \ell_{J4}^{\left(j\right)},\\
\ell_{J2}^{\left(j+1\right)} & = & \ell_{J2}^{\left(j\right)}+\ell_{J5}^{\left(j\right)}+\ell_{h1}^{\left(j\right)},\\
\ell_{J3}^{\left(j+1\right)} & = & \ell_{J2}^{\left(j\right)}+\ell_{J3}^{\left(j\right)}+\ell_{J5}^{\left(j\right)}+2\ell_{h1}^{\left(j\right)},\\
\ell_{J4}^{\left(j+1\right)} & = & \ell_{J3}^{\left(j\right)}+\ell_{J4}^{\left(j\right)}+\ell_{J6}^{\left(j\right)}+\ell_{h1}^{\left(j\right)}+\ell_{h2}^{\left(j\right)},\\
\ell_{J5}^{\left(j+1\right)} & = & \ell_{J2}^{\left(j\right)}+\ell_{J3}^{\left(j\right)}+\ell_{J5}^{\left(j\right)}+\ell_{J6}^{\left(j\right)}+2\ell_{h1}^{\left(j\right)}+\ell_{h2}^{\left(j\right)},\\
\ell_{J6}^{\left(j+1\right)} & = & \ell_{J3}^{\left(j\right)}+\ell_{J4}^{\left(j\right)}+2\ell_{J6}^{\left(j\right)}+\ell_{h1}^{\left(j\right)}+2\ell_{h2}^{\left(j\right)},
\end{array}\label{eq:leffrs4}
\end{equation}
with analogous expressions obtained by interchanging the labels $J$
and $h$. Here, $l_{Ji}^{\left(j\right)}$ or $l_{hi}^{\left(j\right)}$
represent the length of the effective bond $J_{i}$ or field $h_{i}$,
$i\in\left\{ 1,2,\dots,6\right\} $, at the $j$th iteration of the
RG transformation.

The recursion relations in Eqs. \ref{eq:jeffrs4} and their dual counterparts
are valid as long as $h_{1}^{\left(j\right)}>J_{2}^{\left(j\right)}$
and $J_{1}^{\left(j\right)}>h_{2}^{\left(j\right)}$. The first condition
eventually ceases to be valid in the ferromagnetic phase, while the
second one fails in the paramagnetic phase. To be concrete, let us
focus on the ferromagnetic phase. From the dual symmetry of the recursion
relations (under the interchange of bonds and fields), similar results
are valid in the paramagnetic phase.

As in the previous section, it is convenient to introduce a coupling
ratio
\begin{equation}
s_{1}^{\left(j\right)}\equiv\frac{J_{2}^{\left(j\right)}}{h_{1}^{\left(j\right)}},\label{eq:s1rs4}
\end{equation}
whose value allows us to check the validity of the condition $h_{1}^{\left(j\right)}>J_{2}^{\left(j\right)}$.
In order to obtain a closed set of recursion relations, it is also
necessary to define the additional ratios
\begin{equation}
s_{4}^{\left(j\right)}\equiv\frac{J_{5}^{\left(j\right)}}{h_{4}^{\left(j\right)}},\qquad\mbox{and}\qquad r_{i}^{\left(j\right)}\equiv\frac{h_{i}^{\left(j\right)}}{J_{i}^{\left(j\right)}},\ i\in\left\{ 1,2,\dots,6\right\} .\label{eq:rirs4}
\end{equation}
Substitution of Eqs. \ref{eq:jeffrs4} into the above expressions,
and definition of the ratio
\begin{equation}
t^{\left(j\right)}\equiv s_{1}^{\left(j\right)}s_{4}^{\left(j\right)}=s_{1}^{\left(j+1\right)},\label{eq:trs4}
\end{equation}
lead to a set of recursion relations which in matrix form can be written
as
\[
\left(\begin{array}{c}
\ln r_{1}^{\left(j+1\right)}\\
\ln r_{2}^{\left(j+1\right)}\\
\ln r_{3}^{\left(j+1\right)}\\
\ln r_{4}^{\left(j+1\right)}\\
\ln r_{5}^{\left(j+1\right)}\\
\ln r_{6}^{\left(j+1\right)}\\
\ln t^{\left(j+1\right)}
\end{array}\right)=\left(\begin{array}{rrrrrrr}
0 & 0 & 0 & 1 & 0 & 0 & 0\\
1 & 1 & 0 & 0 & 1 & 0 & 0\\
2 & 1 & 1 & 0 & 1 & 0 & 0\\
1 & 1 & 1 & 1 & 0 & 1 & 0\\
2 & 2 & 1 & 0 & 1 & 1 & 0\\
1 & 2 & 1 & 1 & 0 & 2 & 0\\
-1 & -1 & -1 & 0 & 0 & -1 & 2
\end{array}\right)\left(\begin{array}{c}
\ln r_{1}^{\left(j\right)}\\
\ln r_{2}^{\left(j\right)}\\
\ln r_{3}^{\left(j\right)}\\
\ln r_{4}^{\left(j\right)}\\
\ln r_{5}^{\left(j\right)}\\
\ln r_{6}^{\left(j\right)}\\
\ln t{}^{\left(j\right)}
\end{array}\right)
\]
\[
\Rightarrow\ \left|v_{j+1}\right\rangle =\mathbb{T}\left|v_{j}\right\rangle ,
\]
in a notation analogous to that of Eqs. \ref{eq:lnr-recrel} and \ref{eq:vjTk}.
From Eqs. \ref{eq:jnhnrs4}-\ref{eq:trs4}, we see that the initial
value of the vector $\left|v\right\rangle $ takes the form
\[
\left|v_{0}\right\rangle =\left(\begin{array}{c}
2\ln\left(1-\delta\right)\\
2\ln\left(1-\delta\right)\\
4\ln\left(1-\delta\right)\\
6\ln\left(1-\delta\right)\\
6\ln\left(1-\delta\right)\\
8\ln\left(1-\delta\right)\\
2\ln\rho-8\ln\left(1-\delta\right)
\end{array}\right).
\]

The eigenvalues of the matrix $\mathbb{T}$ are $\lambda_{1}=4$,
$\lambda_{2}=2$, $\lambda_{3}=\lambda_{4}=1$, and $\lambda_{5}=\lambda_{6}=\lambda_{7}=0$.
Denoting the respective right eigenvectors by $\left|\phi_{i}\right\rangle $,
$i\in\left\{ 1,2,\dots,7\right\} $, $\left|v_{0}\right\rangle $
can be expanded as
\[
\left|v_{0}\right\rangle =-8\ln\left(1-\delta\right)\left|\phi_{1}\right\rangle +2\ln\rho\left|\phi_{2}\right\rangle -\frac{2}{3}\ln\left(1-\delta\right)\left|\phi_{3}\right\rangle +\left|w_{0}\right\rangle ,
\]
where $\left|w_{0}\right\rangle $ belongs to the kernel of $\mathbb{T}$.
Thus, from $\left|v_{j}\right\rangle =\mathbb{T}^{j}\left|v_{0}\right\rangle $
and from explicit expressions for $\left|\phi_{1}\right\rangle $,
$\left|\phi_{2}\right\rangle $ and $\left|\phi_{3}\right\rangle $,
we obtain, for $i\in\left\{ 1,2,\dots6\right\} $, 
\[
\left\{ \begin{array}{ccl}
\ln r_{i}^{\left(j\right)} & = & \left(f_{i}\lambda_{1}^{j}+g_{i}\lambda_{2}^{j}\right)\ln\left(1-\delta\right),\\
\ln t^{\left(j\right)} & = & -8\lambda_{1}^{j}\ln\left(1-\delta\right)+2\lambda_{2}^{2}\ln\rho,
\end{array}\right.
\]
with the constants $f_{i}$ and $g_{i}$ all positive. It is then
clear that at the critical point $(\delta=0$) we have $r_{i}^{(j)}=1$,
i.e., $h_{i}^{\left(j\right)}=J_{i}^{\left(j\right)}$ at all RG steps,
whereas $t^{(j)}$ asymptotically approaches zero. This indicates
that the RG scheme becomes asymptotically exact, in agreement with
the relevant nature of aperiodic modulations characterized by a wandering
exponent $\omega=\frac{1}{2}>0$. Accordingly, the dynamic scaling
can be shown to assume the activated form
\[
\Omega_{j}\sim\exp\left(-\sqrt{\ell_{j}/\ell_{\rho}}\right),
\]
with $\ell_{j}$ a characteristic length scale and $\ell_{\rho}$
a function of $\rho$ only. This is exactly the form obeyed by the
dynamic scaling behavior of the critical random quantum Ising chain
\cite{fisher92,fisher95}.

In the ferromagnetic phase ($\delta>0$), we have $r_{i}^{\left(j\right)}<1$,
as expected, but the RG scheme is no longer valid once $s_{1}^{\left(j\right)}$
(or equivalently $t^{\left(j\right)}$; see Eq. \ref{eq:trs4}) reaches
unity. This happens for
\begin{equation}
\ln t^{\left(j^{*}\right)}=0\ \Rightarrow\ j^{*}=\frac{\ln\left[\frac{4\ln\left(1-\delta\right)}{\ln\rho}\right]}{\ln\frac{\lambda_{2}}{\lambda_{1}}}\sim\frac{\ln\delta}{\ln\left(\lambda_{2}/\lambda_{1}\right)}.\label{eq:jstarrs4}
\end{equation}
From this step on we need to redefine the RG transformation, since
now $J_{2}^{\left(j\right)}>h_{1}^{\left(j\right)}$. It turns out
that after one step of this new RG transformation all effective fields
are smaller than all effective bonds. Thus, critical properties can
be estimated, in the same way as in the previous section, by looking
at effective quantities after $j_{\delta}\sim j^{*}$ iterations
of the RG scheme.

The matrix connecting effective lengths in consecutive RG steps, built
from Eqs. \ref{eq:leffrs4} and their dual counterparts, has as its
largest eigenvalue $\lambda_{1}=4$, so that the correlation length
can be estimated as
\[
\xi\sim\ell_{j_{\delta}}\sim\lambda_{1}^{j_{\delta}}\sim\delta^{-\nu},
\]
with a correlation length exponent
\[
\nu=\frac{\ln\lambda_{1}}{\ln\lambda_{1}-\ln\lambda_{2}}=2=\frac{1}{1-\omega},
\]
again saturating the Harris-Luck criterion as written in Eq. \ref{eq:invertHL}.

From the recursion relations of the effective magnetic moments, Eq.
\ref{eq:mueffrs4}, we similarly extract a characteristic magnetic
moment scaling as
\[
\mu_{j}\sim\vartheta_{1}^{j},
\]
in which $\vartheta_{1}=3.24698\dots$ is the largest eigenvalue of
the matrix connecting $\left\{ \mu_{i}^{\left(j+1\right)}\right\} $
to $\left\{ \mu_{i}^{\left(j\right)}\right\} $. Thus, the spontaneous
(bulk) magnetization should scale as
\[
m_{x}\sim\frac{\mu_{j_{\delta}}}{\ell_{j_{\delta}}}\sim\left(\frac{\vartheta_{1}}{\lambda_{1}}\right)^{j_{\delta}}\sim\delta^{\beta},
\]
with a critical exponent
\[
\beta=\frac{\ln\left(\vartheta_{1}/\lambda_{1}\right)}{\ln\left(\lambda_{2}/\lambda_{1}\right)}=0.300902\dots.
\]

\begin{figure}
\begin{centering}
\includegraphics[width=0.75\textwidth]{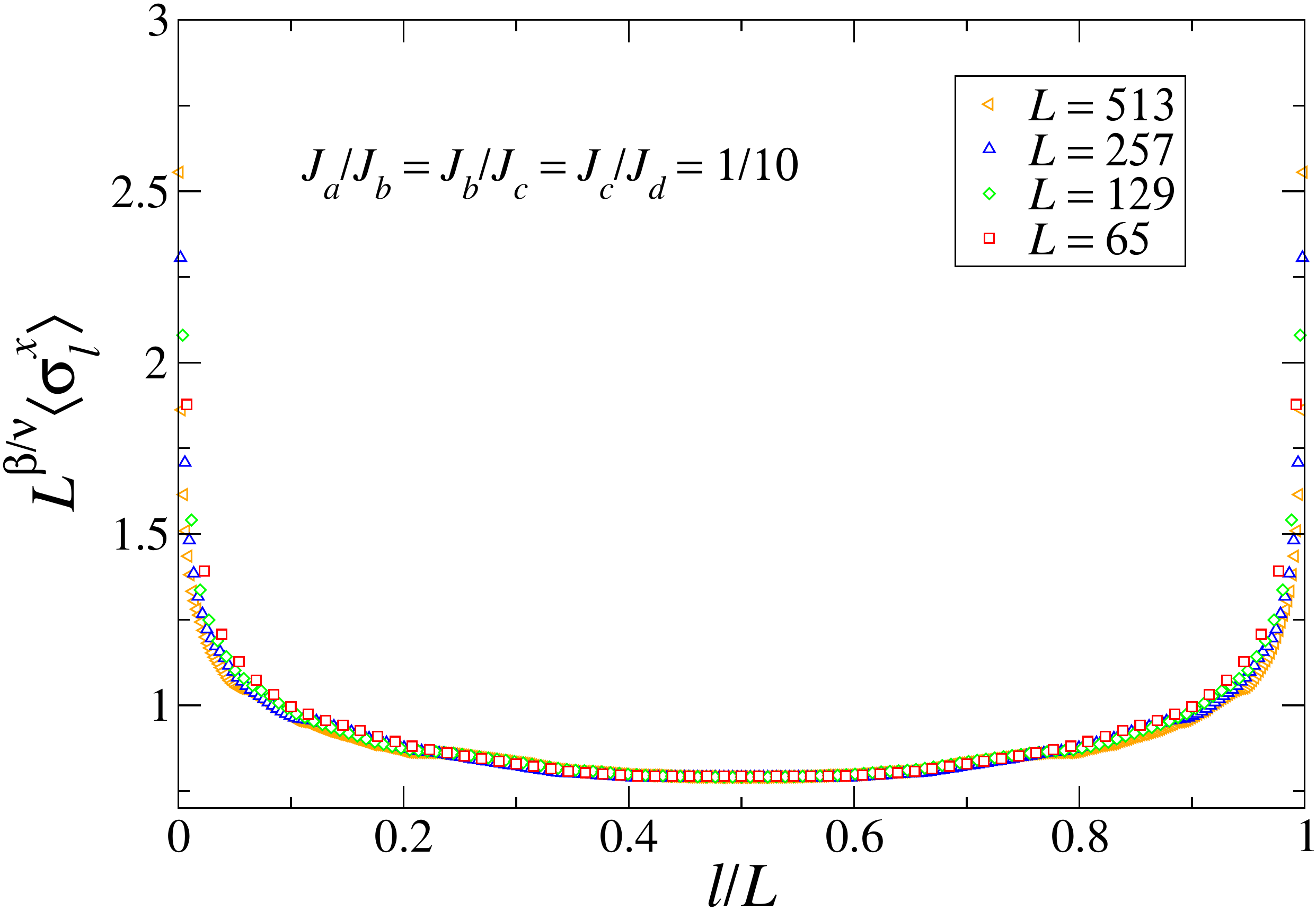}
\par\end{centering}

\caption{\label{fig:maprofrs4}Rescaled average magnetization profiles for the 
four-letter Rudin-Shapiro
sequence, with a coupling ratio $\rho=\frac{1}{10}$.}
\end{figure}
In order to check the validity of the above results, we again resort
to exact diagonalization of finite chains, based on the free-fermion
method. As shown in Fig. \ref{fig:maprofrs4}, these calculations
are fully compatible with the above prediction for the ratio $\beta/\nu$.
Notice that now there is a clear collapse of the curves only for the
bulk magnetization, and the presence of four rather than two distinct
bond leads to the disappearance of the second master curve visible
in Fig. \ref{fig:magprofk3}.

\subsection{The two-letter sequence}

The results derived in the previous subsection are valid as long as
the four bonds assume distinct values. However, by making the identifications
$J_{b}\equiv J_{a}$ and $J_{d}\equiv J_{c}$ we obtain a binary RS
sequence, and the properties of the corresponding quantum Ising chain
can in principle be different. For instance, numerical calculations
and scaling considerations based on a mapping to a directed walk led
Iglói \emph{et al.} \cite{igloi98b} to propose for this binary sequence
a correlation-length critical exponent $\nu=4/3$ and a ratio $\beta/\nu=0.160\left(5\right)$.

\begin{figure}
\begin{centering}
\includegraphics[angle=90,height=0.7\textheight]{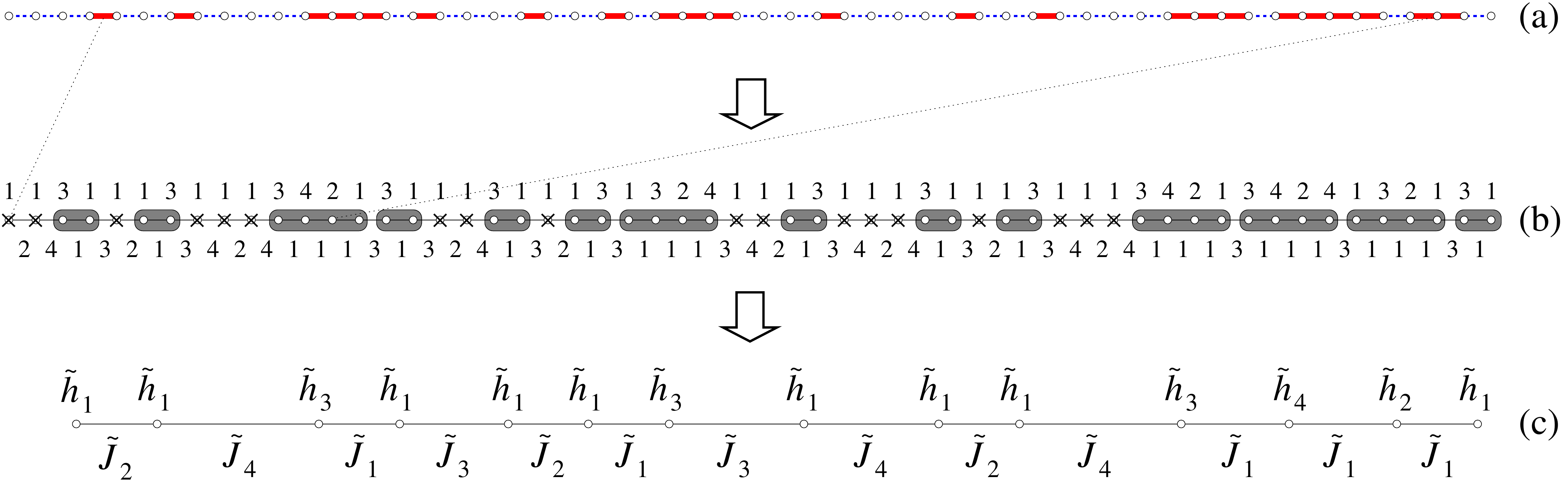}
\par\end{centering}

\caption{\label{fig:rs2sdrg}Renormalization-group transformation as applied
to the two-letter Rudin-Shapiro chain. The original chain is shown
in (a): $J_{a}$ bonds are represented by dashed lines (blue), while
stronger $J_{c}$ bonds appear as solid lines (red); open circles
mark the position of the spins, under a uniform field $h\simeq h_{\mathrm{crit}}$.
Spins connected by $J_{c}$ bonds form clusters, while spins connected
only by $J_{a}$ are decimated. These processes give rise to the effective
chain in (b), characterized by 4 different values of both bonds and
fields, respectively indicated by the numbers below and above the
chain. Upon further lowering of the energy scale, effective spins
under the grey boxes cluster together, while crossed circles
denote effective spins decimated by the action of strong effective
fields. As shown in (c) the next effective chain preserves the structure
in (b). }

\end{figure}
For the quantum Ising chain with couplings following the binary RS
sequence, with $J_{a}=\rho J_{c}$, $0<\rho<1$, the critical field
is given by
\[
h_{\mathrm{crit}}=\rho^{\frac{1}{2}}J_{c}.
\]
When applying the SDRG scheme, we thus focus on the condition $J_{a}\ll h\ll J_{c}$.
As shown in Fig. \ref{fig:rs2sdrg}(a), the original chain has clusters
with $1$ to $4$ strong bonds. At energy scales $\Omega$ such that
$J_{a}<h<\Omega<J_{c}$, this gives rise to $4$ different effective
magnetic moments $\left\{ \mu_{n}\right\} $ and fields $\left\{ h_{n}\right\} $,
with
\begin{equation}
\mu_{n}=\left(n+1\right)\mu\quad\mbox{and}\quad h_{n}=\frac{h^{n+1}}{J_{c}^{n}},\qquad n\in\left\{ 1,2,3,4\right\} .\label{eq:hnrs2}
\end{equation}
At this point, all bonds are equal to $J_{a}$, and the largest energy
scale in the chain is provided by the original fields $h$. Thus,
at energy scales such that $J_{a}<\Omega<h$, all spins under the
action of $h$ can be decimated, inducing $4$ different effective
bonds
\begin{equation}
J_{n}=\frac{J_{a}^{n}}{h^{n-1}},\quad n\in\left\{ 1,2,3,4\right\} .\label{eq:jnrs2}
\end{equation}
The resulting effective chain is depicted in Fig. \ref{fig:rs2sdrg}(b),
and corresponds to a sequence of $4$ different bonds and fields. 

Exactly at the critical point, the bonds and fields are ordered such
that
\[
J_{1}=h_{1}>J_{2}=h_{2}>J_{3}=h_{3}>J_{4}=h_{4},
\]
with 
\[
J_{n}=\rho^{\frac{n+1}{2}}J_{c}.
\]
Notice from Fig. \ref{fig:rs2sdrg}(b) that now, contrary to what
happens for the four-letter RS chain, there are clusters of $1$ to
$3$ strong bonds $J_{1}$ or fields $h_{1}$, and in some of these
clusters both strong bonds and strong fields are present. Of course,
this picture remains valid in the neighborhood of the critical point,
making it harder to write recursion relations, since the perturbative
treatment becomes more involved. However, we can take advantage of
a mapping between the critical quantum Ising chain and the quantum
\emph{XX} chain \cite{fisher94} and of the results derived for the
aperiodic quantum \emph{XX} chain (see Ref. \cite{vieira05b}) to
write the bond recursion relations
\begin{equation}
\begin{array}{lll}
J_{1}^{\left(j+1\right)} & = & f\left(J_{1}^{\left(j\right)},h_{1}^{\left(j\right)}\right)J_{1}^{\left(j\right)}J_{3}^{\left(j\right)},\\
J_{2}^{\left(j+1\right)} & = & \left[f\left(J_{1}^{\left(j\right)},h_{1}^{\left(j\right)}\right)\right]^{2}\left[J_{1}^{\left(j\right)}\right]^{2}J_{2}^{\left(j\right)}J_{3}^{\left(j\right)}/h_{1}^{\left(j\right)},\\
J_{3}^{\left(j+1\right)} & = & f\left(J_{1}^{\left(j\right)},h_{1}^{\left(j\right)}\right)J_{1}^{\left(j\right)}J_{2}^{\left(j\right)}J_{3}^{\left(j\right)}J_{4}^{\left(j\right)}/\left[h_{1}^{\left(j\right)}\right]^{2},\\
J_{4}^{\left(j+1\right)} & = & J_{2}^{\left(j\right)}J_{3}^{\left(j\right)}\left[J_{4}^{\left(j\right)}\right]^{2}/\left[h_{1}^{\left(j\right)}\right]^{3}
\end{array}\label{eq:jeffrs2}
\end{equation}
with dual recursion relations for the effective fields, the function
$f\left(J,h\right)$ being explicitly given by
\[
f\left(J,h\right)=\left(J^{2}+h^{2}\right)^{-\frac{1}{2}}.
\]
The above recursion relations define an RG transformation which maps
the effective chain onto itself as long as 
\begin{equation}
J_{1}^{\left(j\right)}>h_{2}^{\left(j\right)}\quad\mbox{and}\quad h_{1}^{\left(j\right)}>J_{2}^{\left(j\right)}.\label{eq:h1j2rs2}
\end{equation}
Recursion relations for the lengths of the effective couplings take
the form
\begin{equation}
\begin{array}{lll}
\ell_{J1}^{\left(j+1\right)} & = & \ell_{J3}^{\left(j\right)},\\
\ell_{J2}^{\left(j+1\right)} & = & \ell_{J2}^{\left(j\right)}+\ell_{J3}^{\left(j\right)}+\ell_{h1}^{\left(j\right)},\\
\ell_{J3}^{\left(j+1\right)} & = & \ell_{J2}^{\left(j\right)}+\ell_{J3}^{\left(j\right)}+\ell_{J4}^{\left(j\right)}+2\ell_{h1}^{\left(j\right)},\\
\ell_{J4}^{\left(j+1\right)} & = & \ell_{J2}^{\left(j\right)}+\ell_{J3}^{\left(j\right)}+2\ell_{J4}^{\left(j\right)}+3\ell_{h1}^{\left(j\right)},\\
\ell_{h1}^{\left(j+1\right)} & = & \ell_{h1}^{\left(j\right)}+\ell_{h3}^{\left(j\right)}+\ell_{J1}^{\left(j\right)},\\
\ell_{h2}^{\left(j+1\right)} & = & 2\ell_{h1}^{\left(j\right)}+\ell_{h2}^{\left(j\right)}+\ell_{h3}^{\left(j\right)}+3\ell_{J1}^{\left(j\right)},\\
\ell_{h3}^{\left(j+1\right)} & = & \ell_{h1}^{\left(j\right)}+\ell_{h2}^{\left(j\right)}+\ell_{h3}^{\left(j\right)}+\ell_{h4}^{\left(j\right)}+3\ell_{J1}^{\left(j\right)},\\
\ell_{h4}^{\left(j+1\right)} & = & \ell_{h2}^{\left(j\right)}+\ell_{h3}^{\left(j\right)}+2\ell_{h4}^{\left(j\right)}+3\ell_{J1}^{\left(j\right)},
\end{array}\label{eq:leffrs2}
\end{equation}
with the initial values $\ell_{J1}^{\left(0\right)}=\frac{1}{2}$,
$\ell_{J2}^{\left(0\right)}=\ell_{h1}^{\left(0\right)}=\frac{3}{2}$,
$\ell_{J3}^{\left(0\right)}=\ell_{h2}^{\left(0\right)}=\frac{5}{2}$,
$\ell_{J4}^{\left(0\right)}=\ell_{h3}^{\left(0\right)}=\frac{7}{2}$,
$\ell_{h4}^{\left(0\right)}=\frac{9}{2}$. 

At the critical point, $f\left(J_{1}^{\left(j\right)},h_{1}^{\left(j\right)}\right)$
reduces to $f\left(J_{1}^{\left(j\right)},J_{1}^{\left(j\right)}\right)=1/\sqrt{2}J_{1}^{\left(j\right)}$,
and the recursion relations in Eqs. \ref{eq:jeffrs2} recover the
multiplicative structure satisfied by the previously discussed aperiodic
chains. Then, we can show by the method described for the four-letter
RS sequence that again the SDRG approach becomes asymptotically exact,
predicting a dynamic scaling form
\[
\Omega_{j}\sim\exp\left(-\sqrt{\ell_{j}/\ell_{\rho}}\right),
\]
consistent with the value $\omega=\frac{1}{2}$ of the wandering exponent
of the binary RS sequence.

The analytical treatment of the recursion relations in the off-critical
regime is hindered by the presence of the function $f\left(J,h\right)$.
It is possible to perform an expansion on the distance $\delta$
to criticality in order to determine at which step of the RG transformation
the condition in Eq. \ref{eq:h1j2rs2} ceases to be valid. However,
not only is the resulting equation impossible to solve in closed form,
but also the new RG transformation which we have to impose does not
seem to lead to an invariant sequence of effective couplings. Thus,
we have to recourse to a numerical implementation of the SDRG scheme. 

In order to reach the largest possible system sizes, we start at a
given distance to criticality $\delta$ (defined as in Eq. \ref{eq:hcrs4})
and apply the recursion relations \ref{eq:jeffrs2} (and their dual
counterparts) up to the point at which $\min\left\{ J_{1}^{\left(j\right)},h_{1}^{\left(j\right)}\right\} <\max\left\{ J_{2}^{\left(j\right)},h_{2}^{\left(j\right)}\right\} $,
keeping track of the effective lengths via Eqs. \ref{eq:leffrs2}.
Let us denote the effective couplings at this point by $\left\{ J_{n}^{\left(j^{*}\right)}\right\} $
and $\left\{ h_{n}^{\left(j^{*}\right)}\right\} $. Since up to this
point the sequence of effective couplings is invariant under the RG
transformation, it can be generated by applying the rules leading
to Eqs. \ref{eq:hnrs2} and \ref{eq:jnrs2} to a chain with bonds
following the binary RS sequence. To the resulting effective couplings,
we then assign the values $\left\{ J_{n}^{\left(j^{*}\right)}\right\} $
and $\left\{ h_{n}^{\left(j^{*}\right)}\right\} $,  and iterate numerically
the SDRG scheme by sweeping through the lattice and progressively
eliminating the high-energy degrees of freedom. For a binary RS sequence
with $2^{20}\simeq10^{6}$ letters, the corresponding number of $\left\{ J_{n}\right\} $
and $\left\{ h_{n}\right\} $ couplings is around $5\times10^{5}$,
and correcting for the lengths of the effective couplings leads to
a real chain containing $6\times10^{12}$ spins for $\delta=10^{-5}$.

\begin{figure}
\begin{centering}
\includegraphics[width=0.75\textwidth]{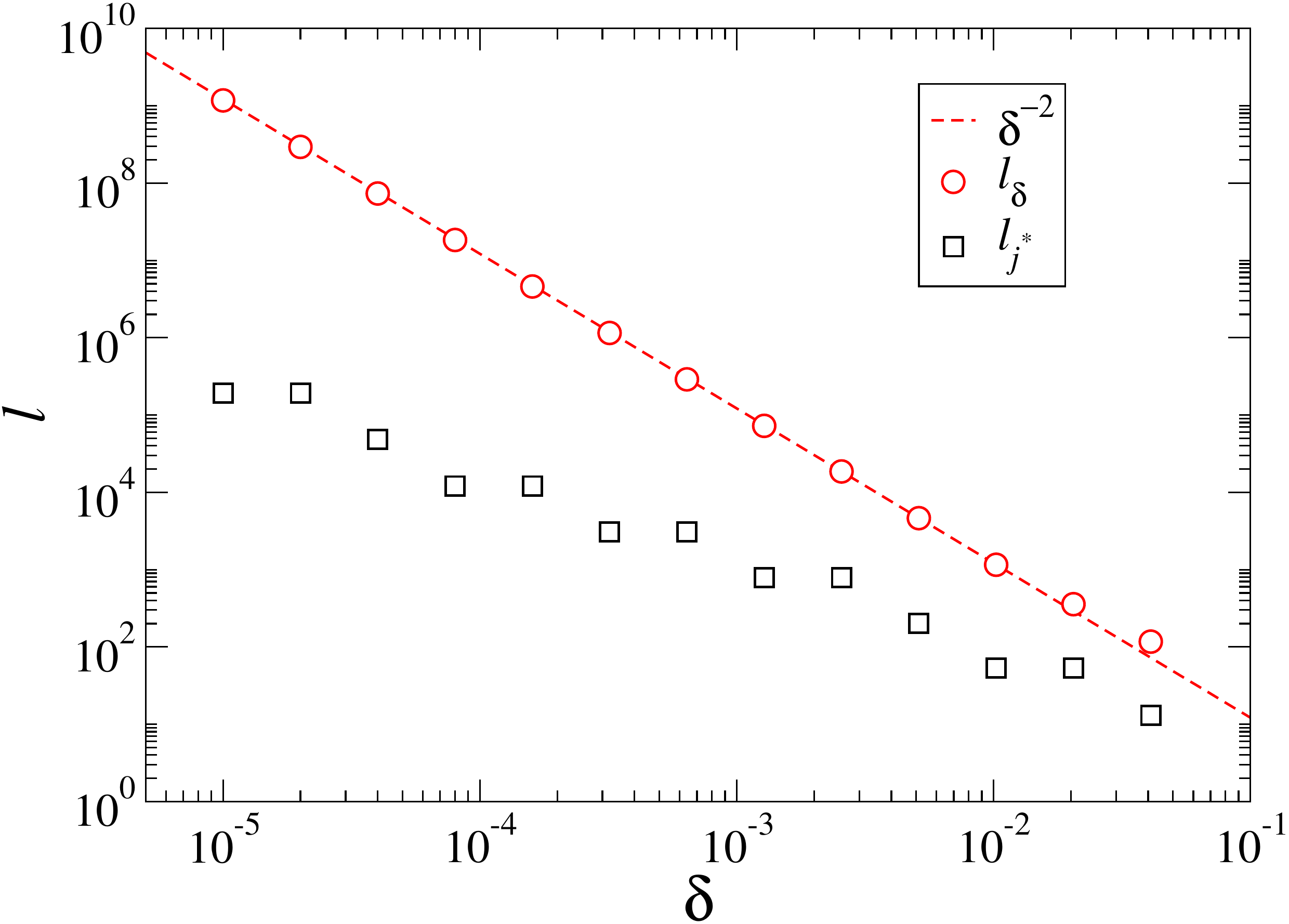}
\par\end{centering}

\caption{\label{fig:lversuseps}Dependence of the lengths $l_{\delta}$
(proportional to the correlation length) and $l_{j^{*}}$ on the distance
to criticality $\delta$, for a coupling ratio $J_{a}/J_{c}$=$\frac{1}{2}$.}

\end{figure}
We implement this numerical procedure, in the paramagnetic phase,
up to the point at which all bonds become smaller than all fields.
In Fig. \ref{fig:lversuseps} we plot, for $\rho=\frac{1}{2}$, the
scaling behavior of the length $\ell_{\delta}$ of the strongest
bond at that point, as a function of the distance to criticality.
Also shown is the length $\ell_{j^{*}}$ of the strongest bond $J_{1}^{\left(j^{*}\right)}$
at the point beyond which the sequence of effective lengths is no
longer invariant. We see that $\ell_{\delta}$ nicely follows a
power law $\ell_{\delta}\sim\left|\delta\right|^{-2}$, and thus
we conclude that, using $\ell_{\delta}$ as an estimate of the correlation
length, we obtain 
\[
\xi\sim\left|\delta\right|^{-\nu},
\]
with an exponent $\nu=2$. 

This result is in contrast to that obtained by Iglói et al. from a
scaling argument for the surface magnetization of finite chains, based
on an equivalence of the problem with a directed walk. As already
mentioned, these authors predict a correlation-length critical exponent
$\nu=\frac{4}{3}$. They also provide an estimate of the ratio $\beta/\nu\simeq0.160$,
by adjusting numerical results for critical magnetization profiles
of small chains (containing between $9$ and $65$ spins), with a
ratio between weak and strong bonds given by $J_{a}/J_{c}=1/16$.
Incidentally, they find the surprising result that the data for such
small chains can be well fitted by the predictions of conformal invariance,
although the system is clearly not conformally invariant, in view
of the strongly anisotropic scaling along the space and time directions.

\begin{figure}
\begin{centering}
\includegraphics[width=0.8\textwidth]{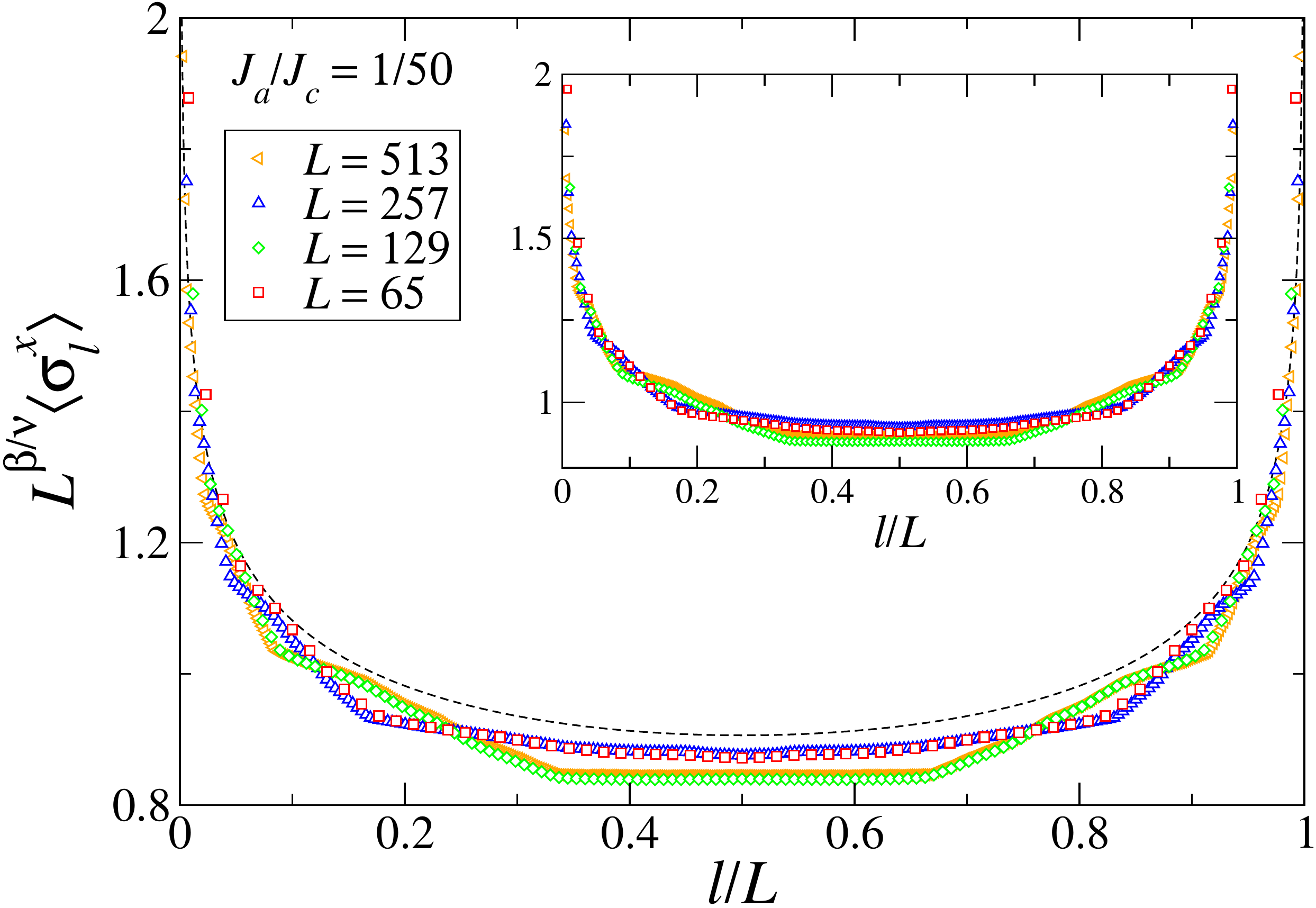}
\par\end{centering}

\caption{\label{fig:magprof-rs2}Rescaled average magnetization profiles for the two-letter
Rudin-Shapiro quantum Ising chain. Main panel: rescaling with the
exponents calculated by the SDRG approach. Inset: rescaling with the
numerical estimate of Ref. \cite{igloi98b}. The dashed line in the
main panel corresponds to the conformal-invariance prediction,
Eq. \ref{eq:mxconfinv}.}

\end{figure}
We then decided to use the free-fermion method to look at larger chain
sizes, and smaller bond ratios, in order to check whether we could
fit the numerical results with our prediction for $\beta/\nu$, assuming,
on the grounds of universality, that the ratio $\beta/\nu=0.150451\dots$
found for the four-letter RS chain would still be valid for the two-letter
RS chain. Figure \ref{fig:magprof-rs2} shows the results of free-fermion
calculations of critical magnetization profiles for chains containing
between $65$ and $513$ spins, and $J_{a}/J_{c}=1/50$. Despite the
small difference between our prediction for $\beta/\nu$ and that
of Ref. \cite{igloi98b}, the inset shows that our prediction yields
a better data collapse. Although the results shown in Fig. \ref{fig:magprof-rs2}
are definitely incompatible with the conformal-invariance form 
\begin{equation}
m_{x}\left(l/L\right)\sim\left[L\sin\left(\pi l/L\right)\right]^{-\beta/\nu},
\label{eq:mxconfinv}
\end{equation}
we do observe a reasonable agreement with such form for small chains
(not shown), as in Ref. \cite{igloi98b}. This agreement becomes poorer
as the coupling ratio is reduced.

\begin{figure}
\begin{centering}
\includegraphics[width=0.8\textwidth]{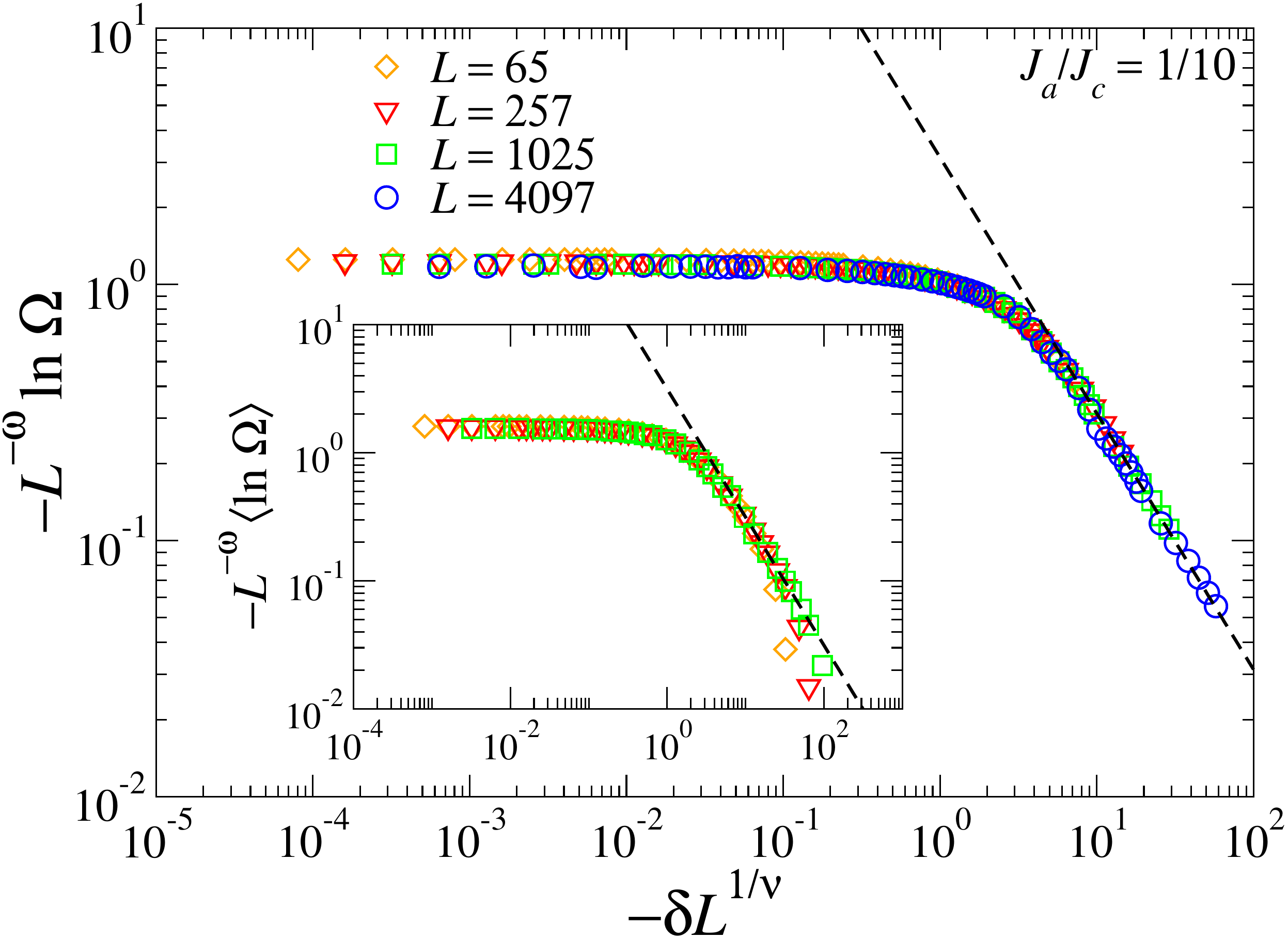}
\par\end{centering}

\caption{\label{fig:fssgap}Finite-size scaling plot of the lowest gap calculated
numerically for off-critical (paramagnetic) quantum Ising chains with couplings following
the binary Rudin-Shapiro sequence. Parameters correspond to $\omega=\frac{1}{2}$
and $\nu=2$. 
The dashed lines are proportional to 
$\left(\left|\delta\right| L^{1\text{/\ensuremath{\nu}}}\right)^{-\omega\nu}.$
Values of $\left|\delta\right|$ range from $10^{-4}$ ($10^{-3}$ for the inset)
to $8\times 10^{-1}$ ($8\times 10^{0}$ for the inset) for all system sizes.
}

\end{figure}
Furthermore, the results in Eqs. \ref{eq:dynrelrelevant} and \ref{eq:gapvseps}
can be used to check directly the prediction $\nu=2$, from free-fermion
calculations of the lowest gap $\Omega_{L}\left(\delta\right)$
for chains with $L$ spins, at a distance $\delta$ from criticality,
via the finite-size scaling \emph{ansatz}
\[
-\ln\Omega_{L}\left(\delta\right)=L^{\omega}f\left(\left|\delta\right| L^{1/\nu}\right),
\]
in which $f\left(x\right)$ is a scaling function predicted to behave
as
\[
f\left(x\right)\sim x^{-\omega\nu}
\]
for $x\gg1$, corresponding to large system sizes at a fixed $\delta$,
and to approach a constant as $x\rightarrow 0$,
which corresponds to $\left|\delta\right|\rightarrow 0$.
As seen in Fig. \ref{fig:fssgap}, we obtain a very good
data collapse for binary Rudin-Shapiro chains with coupling ratio
$J_{a}/J_{c}=1/10$, and lengths ranging from $L-1=64$ to $L-1=4096$,
by setting $\omega=\frac{1}{2}$ and using the SDRG prediction $\nu=2$.
(The calculations were performed in the paramagnetic phase, corresponding
to $\delta<0$, for which the lowest gap is obtained from the smallest
positive fermion energy. In the ferromagnetic phase this vanishes exponentially 
with the system size, and the lowest gap is obtained instead from the second smallest
positive fermion energy.) 
The main plot in Fig. \ref{fig:fssgap} shows the rescaled lowest gap produced
by the subsequence of $L-1$ bonds
($L=65$, $257$, $1025$, $4097$)
obtained, starting from a single letter $a$, by iterating the substitution rule 
in Eq. \ref{eq:rs4subst}, and imposing $J_{b}\equiv J_{a}$ and $J_{d}\equiv J_{c}$.
For the sake of illustration, the inset shows the corresponding results obtained by averaging the 
logarithm of the lowest gaps
for all subsequences of $L-1$ bonds cut from an infinite two-letter Rudin-Shapiro 
sequence. (Due to computational costs, chains with $L=4097$ were not available
for averaging.) Results for the average lowest gap, as well as for the average
second gap, give qualitatively the same scaling picture (not shown),
in agreeement with SDRG predictions.

\section{\label{sec:Discussion-and-conclusions}Discussion and conclusions}

In this paper, we have presented an extension of the strong-disorder
renormalization group approach for the study of the low-energy critical
behavior of the quantum Ising chain with couplings chosen from aperiodic
but deterministic sequences, generated by substitution rules.

For the cases of marginal and relevant aperiodicity, in which the
critical behavior is expected to depart from the Onsager universality
class, we have been able to obtain analytical results for various
critical exponents, such as those related to the correlation length,
the bulk magnetization, and the spin-spin correlation functions at
criticality. Interestingly, we always find that the correlation-length
critical exponent is given by $\nu=(1-\omega)^{-1}$, a result which
barely satisfies the Harris-Luck criterion \cite{luck93b,harris74} 
around the aperiodic
fixed point, which takes the form $\nu\geq(1-\omega)^{-1}$.
We have also derived results for the off-critical lowest
energy gaps, which reflect the absence of Griffiths singularities
for the present class of aperiodic systems, as expected from previous
arguments \cite{igloi98b,hermisson00}. All our results have been
fully confirmed by numerical calculations on finite chains, based
on the free-fermion methods. 

The basic physical picture offered by the SDRG approach, derived for
a family of two-letter substitution rules, has been shown to remain
valid for more complicated sequences, represented by the two- and
four-letter Rudin-Shapiro sequences. According to this picture, 
sufficiently close to the critical
point, as the energy scale is progressively reduced, the effective
chains generated along the RG process exhibit a self-similar structure,
in which the sequence of effective couplings remains invariant and
their values decrease, whereas the magnetic moments of the effective
spins and the distances between those spins grow exponentially. This
picture holds up to a certain step of the RG process, beyond which
the ground state of the chain settles into either the ferromagnetic
or the paramagnetic phases, characterized respectively by the dominance
of bonds or fields. 

The self-similar nature of the effective chains is helpful
in understanding the absence of Griffiths singularities	in the
systems studied here. At all length scales, self-similarity
guarantees that geometric fluctuations are entirely governed by the 
wandering exponent $\omega$, setting a maximum size for 
locally critical clusters, as already noted in \cite{hermisson00}.
This is in contrast with the uncorrelated random quantum Ising
chain, where the wandering exponent only gauges the growth of
average geometric fluctuations, there being a nonzero probability
for the occurrence of arbitrarily large clusters of spins which
are locally in a different phase than the bulk.

Finally, the results reported here, concerning critical exponents
and the nature of the ground-state phases, remain valid for aperiodic
\emph{XY} chains, described by the Hamiltonian
\[
\mathcal{H}=-\frac{1}{2}\sum J_{i}\left[\left(1+\gamma_{i}\right)\sigma_{i}^{x}\sigma_{i+1}^{x}+\left(1-\gamma_{i}\right)\sigma_{i}^{y}\sigma_{i+1}^{y}\right]-\sum_{i}h_{i}\sigma_{i}^{z},
\]
as long as the anisotropy parameters $\gamma_{i}$, 
$0\leq\left|\gamma_{i}\right|\leq 1$, 
are all positive
or all negative. Under these conditions, the SDRG scheme is straightforwardly
adapted, and it can be shown \cite{deoliveira2011} that the anisotropies
$\gamma_{i}$ flow towards $\left|\tilde{\gamma_{i}}\right|=1$ as
the RG scheme is iterated, so that the low-energy behavior corresponds
to that of the quantum Ising chain.

\ack{}{We thank J. A. Hoyos for useful conversations and a critical reading
of the manuscript. This work has been supported by the Brazilian agencies
CNPq and FAPESP.}

\vspace{0.05\paperheight}


\end{document}